\documentclass{pasj02}

\usepackage{amsmath}	
\usepackage[whole]{bxcjkjatype}
\usepackage{natbib}
\usepackage{url}
\usepackage[switch,mathlines]{lineno}
\usepackage{bm}
\usepackage{tikz}
\usetikzlibrary{arrows.meta,positioning,calc}

\newcommand{\nar}{New Astron. Rev.}

\def\arcdeg{$^\circ$}

\Received{2026/06/02}
\Accepted{2026/07/13}
\Published{yyyy/mm/dd}

\begin{document} 

\title{
A differentiable hydrodynamical approach to the Milky Way bar pattern speed with CO longitude--velocity data
}

\author{Junichi \textsc{Baba}\altaffilmark{1,2}\altemailmark \orcid{0000-0002-2154-8740}}
\author{Kohei \textsc{Hattori}\altaffilmark{2,3,4}\altemailmark \orcid{0000-0001-6924-8862}}
\email{babajn2000@gmail.com; junichi.baba@sci.kagoshima-u.ac.jp}

\altaffiltext{1}{Amanogawa Galaxy Astronomy Research Center, Graduate School of Science and Engineering, Kagoshima University, 1-21-35 Korimoto, Kagoshima 890-0065, Japan.}
\altaffiltext{2}{National Astronomical Observatory of Japan, Mitaka, Tokyo 181-8588, Japan.}
\altaffiltext{3}{The Graduate University for Advanced Studies, SOKENDAI, 2-21-1 Osawa, Mitaka, Tokyo 181-8588, Japan.}
\altaffiltext{4}{The Institute of Statistical Mathematics, 10-3 Midoricho, Tachikawa, Tokyo 190-8562, Japan.}


\maketitle

\begin{abstract}
We present a differentiable hydrodynamical framework for modeling barred gas flow in the Milky Way and constraining broad low-loss regions in the bar-pattern-speed parameter space from Galactic gas longitude--velocity data.
The method evolves a neutral-gas disk in a fixed barred potential for a given gas response time, projects it into longitude--velocity ($\ell$--$v$) space, and compares predicted and target maps using a loss based on the cosine similarity of processed and masked $\ell$--$v$ maps.
This loss emphasizes large-scale morphology rather than the absolute gas-emission scale.
Because the forward model is differentiable, we can compute gradients with respect to the bar pattern speed and optimize the model directly in observable space.
We validate the method with self-consistency hydrodynamical mocks, in which the target and fitting maps are generated with the same differentiable solver, and with an independent hydrodynamical mock that includes more realistic interstellar-medium physics.
These tests recover or identify low-loss regions near the input pattern speed, showing that the method captures coherent bar-driven structures in $\ell$--$v$ space.
We then apply the method to the observed CO $\ell$--$v$ structure of the inner Milky Way.
For the observed CO data, we identify broad low-loss regions rather than defining the pattern speed from a single optimization run.
These regions include moderate pattern speeds, $|\Omega_{\rm b}|\sim30$--$40\,{\rm km\,s^{-1}\,kpc^{-1}}$, consistent with current stellar-dynamical constraints.
Their exact location depends on modeling choices such as the gas response time and viewing angle.
Thus, the present analysis does not determine a single precise value of $\Omega_{\rm b}$.
This first application demonstrates the feasibility of differentiable hydrodynamical modeling of Galactic gas as an independent kinematic test of barred Milky Way models and opens a path toward multi-parameter forward modeling of Galactic gas data in position--position--velocity space.
\end{abstract}


\section{Introduction}
\label{sec:introduction}

The Milky Way bar is one of the most important non-axisymmetric structures in the Galaxy \citep[][]{Bland-HawthornGerhard2016}.
It redistributes angular momentum, shapes stellar and gas orbits in the inner disk, and connects the bulge, the disk, and the central molecular zone \citep[e.g.,][]{FriedliBenz1995,Seo+2019,BabaKawata2020a}.
Its dynamical impact is controlled by its mass distribution, strength, pattern speed, and orientation \citep[e.g.,][]{Athanassoula1992b,Wada+1994,Sormani+2015c}.
Among these quantities, the bar pattern speed $\Omega_{\rm b}$ and the bar viewing angle $\phi_{\rm b}$ are especially important for connecting dynamical models of the inner Milky Way to observations.
The pattern speed sets the resonance structure and the time dependence of the rotating non-axisymmetric potential, while the viewing angle determines how the bar is seen from the Solar position.

The Gaia mission has transformed the study of Milky Way dynamics \citep[][]{Perryman2026}.
Together with spectroscopic and photometric surveys, Gaia has enabled detailed measurements of stellar density, kinematics, phase-space substructure, and disequilibrium features over a large fraction of the Galactic disk.
These data have been used to constrain the Galactic bar through stellar orbits, moving groups, resonant signatures, phase-space ridges, and global dynamical models \citep[e.g.,][]{Portail+2017,Monari+2019,Sanders+2019,Asano+2020,Binney2020,ChibaSchoenrich2021,ClarkeGerhard2022}.
Recent studies generally favor a moderately inclined bar, with a bar viewing angle of roughly $\phi_{\rm b} \simeq 25^{\circ} \pm 10^{\circ}$ and a relatively slow pattern speed of about $|\Omega_{\rm b}| \simeq 30$--$40\,{\rm km\,s^{-1}\,kpc^{-1}}$, although the inferred values remain method dependent \citep[see the review by][]{HuntVasiliev2025}.
An important next step is therefore not only to refine these stellar-dynamical constraints, but also to test whether they are consistent with independent tracers of the same barred potential.

Cold interstellar gas provides such an independent tracer.
Because gas is dissipative, it responds to the rotating bar through shocks, offset dust-lane flows, orbit crowding, nuclear-ring formation, and non-circular streaming motions \citep[e.g.,][]{Athanassoula1992b}.
These motions are projected into H,{\sc i} and CO longitude--velocity ($\ell$--$v$) diagrams, where $\ell$ denotes Galactic longitude and $v$ denotes the line-of-sight velocity relative to the local standard of rest \citep[e.g.,][]{Dame+2001,Kalberla+2005}.
Although an $\ell$--$v$ diagram does not provide a unique face-on gas map, it preserves coherent line-of-sight velocity structures.
In a barred disk, radial and azimuthal streaming motions move gas away from the circular-rotation locus and generate high-velocity ridges, forbidden-velocity emission, and asymmetric high-velocity features \citep[e.g.,][]{Binney+1991,Sormani+2015c}.
The positions and velocities of these structures depend on the gas response to the rotating bar, and therefore on $\Omega_{\rm b}$ and $\phi_{\rm b}$ \citep[e.g.,][]{Bissantz+2003}.

Previous studies have compared observed H,{\sc i} and CO $\ell$--$v$ diagrams with barred gas-flow models to constrain the bar viewing angle, pattern speed, and related parameters \citep[e.g.,][]{MulderLiem1986,Wada+1994,WeinerSellwood1999,Fux1999,EnglmaierGerhard1999,Bissantz+2003,Rodriguez-FernandezCombes2008,Baba+2010,Pettitt+2014,Sormani+2015c,Li+2016,Li+2022}.
Several of these studies made quantitative comparisons using terminal-velocity envelopes, selected $\ell$--$v$ features, or feature-based diagnostics.
In particular, \citet{SormaniMagorrian2015} introduced an automated feature-matching method based on broad $\ell$--$v$ structures, while \citet{Li+2022} quantified the mismatch of selected physical features such as the 3 kpc arms, forbidden-velocity region, and terminal-velocity curve.
However, these approaches did not propagate gradients from the $\ell$--$v$ mismatch back to bar parameters through the full gas-dynamical calculation.

To enable such gradient-based inference, we formulate the problem as a differentiable forward-modeling problem.
In forward modeling, one starts from a physical model, predicts observables through a forward calculation, and compares those predictions directly with the data.
This is useful when the observable depends on the physical parameters through a complex and non-linear process, making direct analytic inversion impractical.
This perspective is closely related to simulation-based inference and physics-informed modeling \citep[e.g.,][]{Cranmer+2020,Ting2026ARAA}.
It is also related to differentiable physical simulation \citep[e.g.,][]{Baydin+2018}, where ``differentiable'' means that gradients of the objective function with respect to the model parameters can be propagated automatically through the full forward calculation, without requiring finite-difference approximations.
In the present problem, a change in the bar parameters changes the gas response, the projected $\ell$--$v$ diagram, and the resulting mismatch between the model and the data.
These gradients provide information for direct optimization and a scalable route to higher-dimensional parameter inference.

In this paper, we apply this idea to the CO $\ell$--$v$ diagram of the inner Milky Way.
We adopt a fixed barred Milky Way potential based on the stellar-dynamically constrained model of \citet{Portail+2017}, using its analytic representation by \citet{Sormani+2022agama}.
This choice fixes the mass distribution and strength of the bar.
We compute the gas response with a non-self-gravitating, isothermal hydrodynamical model and compare the projected $\ell$--$v$ structure with observed CO data.
The parameters that most directly affect the projected gas kinematics are the bar pattern speed $\Omega_{\rm b}$, the viewing angle $\phi_{\rm b}$, and the gas response time $t_{\rm end}$ (the duration of the hydrodynamical evolution).
In the present study, we focus mainly on $\Omega_{\rm b}$, while treating $\phi_{\rm b}$ and $t_{\rm end}$ as nuisance parameters.
As a first step, we ask whether gas $\ell$--$v$ diagrams contain enough information to recover $\Omega_{\rm b}$ in controlled experiments, and whether the same framework gives physically meaningful low-loss regions when applied to the observed CO emission of the Milky Way.

This paper is organized as follows.
In Section~\ref{sec:methods}, we describe the differentiable barred-gas forward model, including the hydrodynamical solver, observer projection, soft longitude--velocity binning, and objective function.
In Section~\ref{sec:mock_validation}, we validate the method using both self-consistency hydrodynamical mock data and independent smoothed-particle hydrodynamics (SPH) mock data, and examine the main parameter degeneracies and robustness checks.
In Section~\ref{sec:data_application}, we apply the framework to observed CO $\ell$--$v$ data, perform gradient-based fits, and examine whether the resulting low-loss regions are consistent with stellar-dynamical constraints on the Milky Way bar.
We discuss the implications, limitations, and future directions in Section~\ref{sec:discussion}.

\section{Differentiable barred-gas forward model}
\label{sec:methods}

In this section, we describe the differentiable barred-gas forward model used to connect the bar parameters to observed gas $\ell$--$v$ diagrams.
We first define the differentiable formulation in Section~\ref{subsec:differentiable_modeling}.
We then describe the fixed barred Milky Way potential and the isothermal hydrodynamics in Sections~\ref{subsec:potential} and \ref{subsec:hydro}.
Next, we describe the differentiable projection from the simulated gas disk to observable $\ell$--$v$ space in Section~\ref{subsec:lv_projection}.
Finally, we define the objective function and describe the optimization strategy in Section~\ref{subsec:objective_optimization}.

\subsection{Differentiable modeling}
\label{subsec:differentiable_modeling}

In this work, ``differentiable'' means that the predicted $\ell$--$v$ map varies smoothly with the model parameters, so that the sensitivity of the final objective function to those parameters can be computed by automatic differentiation.
Let $\Theta$ denote the set of model parameters.
In the present application, the main optimized parameter is the bar pattern speed $\Omega_{\rm b}$.
The bar viewing angle $\phi_{\rm b}$ and the gas response time $t_{\rm end}$ also affect the projected $\ell$--$v$ structure and are treated as nuisance parameters below.

The forward model first defines a mapping to a raw projected $\ell$--$v$ map,
\begin{equation}
    \hat{I}_{\rm raw}(\ell,v)
    =
    {\cal F}_{\rm raw}(\Theta),
    \label{eq:forward_operator}
\end{equation}
where $\hat{I}_{\rm raw}(\ell,v)$ is the raw predicted $\ell$--$v$ map before the processing step, and ${\cal F}_{\rm raw}$ denotes the gas-dynamical evolution (Section~\ref{subsec:hydro}) followed by the projection to observable space (Section~\ref{subsec:lv_projection}).
The map used in the objective function is then obtained by applying a processing operator ${\cal P}$,
\begin{equation}
    \hat{I}(\ell,v)
    =
    {\cal P}\left[\hat{I}_{\rm raw}(\ell,v)\right],
    \label{eq:processing_operator}
\end{equation}
where ${\cal P}$ denotes the map processing defined in Section~\ref{subsec:objective_optimization}, including logarithmic compression and robust percentile-based normalization of the map intensities.

We compare the predicted map with a target map through an objective function,
\begin{equation}
    \mathcal{L}
    =
    \mathcal{L}\left[\hat{I}(\ell,v), I^{\rm target}(\ell,v)\right],
    \label{eq:objective_operator}
\end{equation}
whose explicit form is given in Section~\ref{subsec:objective_optimization}.
Because the numerical operations in the forward calculation and in the processing operator are differentiable, the gradient
\begin{equation}
    \frac{\partial \mathcal{L}}{\partial \Theta}
\end{equation}
can be computed by automatic differentiation.
We implement the full forward calculation in {\tt PyTorch} \citep{Paszke+2019}, so that gradients can be propagated from the objective function through the map processing, the $\ell$--$v$ projection, and the hydrodynamical time integration back to the model parameters.
We use these gradients for optimization and for diagnostic tests of the loss behavior.

\begin{figure*}
\begin{center}
\includegraphics[width=0.95\textwidth]{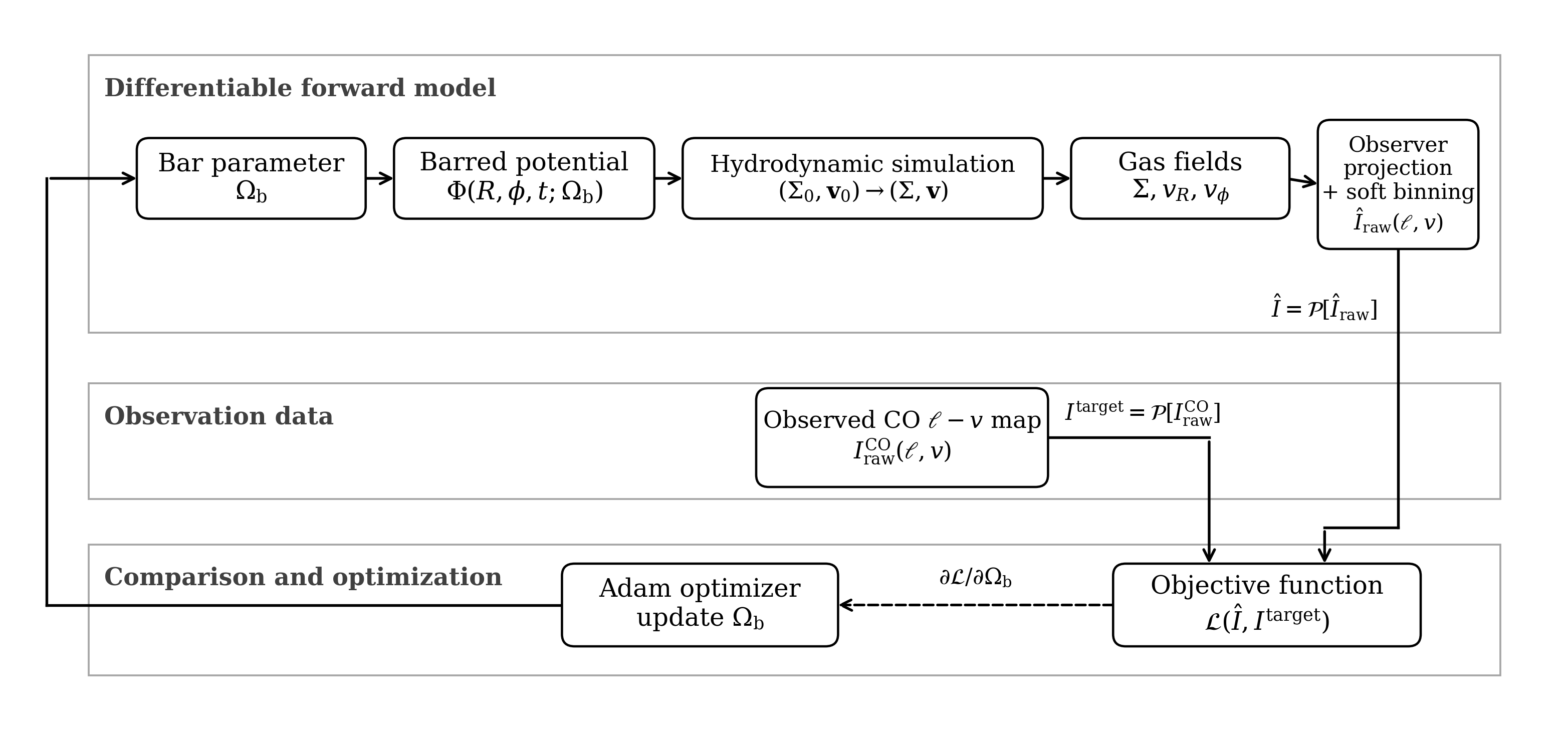}
\end{center}
\caption{
Schematic overview of the differentiable barred-gas forward model.
The top panel shows the forward model.
The bar pattern speed $\Omega_{\rm b}$ defines the rotating barred potential, and the hydrodynamic simulation evolves the initial gas state $(\Sigma_0,\boldsymbol{v}_0)$ into the gas fields $(\Sigma,v_R,v_\phi)$.
These gas fields are then mapped into the observable $\ell$--$v$ space through the observer projection and soft-binning operator, producing the raw model map $\hat{I}_{\rm raw}(\ell,v)$.
The label $\hat{I}={\cal P}[\hat{I}_{\rm raw}]$ on the downward arrow indicates the processing step that converts the raw model map into the processed model map used by the objective function.
This processing includes logarithmic compression and robust percentile-based normalization of the map intensities.
The middle panel shows the raw observed CO map, $I_{\rm raw}^{\rm CO}(\ell,v)$.
The label $I^{\rm target}={\cal P}[I_{\rm raw}^{\rm CO}]$ indicates the processing step that converts the raw observed CO map into the processed target map.
The bottom panel shows the comparison and optimization step.
The objective function $\mathcal{L}(\hat{I},I^{\rm target})$ measures the mismatch between the processed model and target maps.
Automatic differentiation gives the gradient $\partial\mathcal{L}/\partial\Omega_{\rm b}$, which is passed to the Adam optimizer to update $\Omega_{\rm b}$.
The gas fields are intermediate outputs of the hydrodynamic calculation and are not optimized as independent parameters.
\textbf{Alt text:} Flowchart showing how the bar pattern speed is used to evolve a gas disk, project the result into longitude--velocity space, process the model and observed CO maps in the same way, compute a loss, and update the pattern speed with automatic differentiation and an optimizer.
}
\label{fig:gradient_propagation_chart}
\end{figure*}

The forward model can be summarized as
\begin{equation}
    (\Omega_{\rm b}, \phi_{\rm b}, t_{\rm end})
    \longrightarrow
    \left(
    \Sigma, v_R, v_\phi
    \right)
    \longrightarrow
    \hat{I}_{\rm raw}(\ell,v)
    \longrightarrow
    \hat{I}(\ell,v)
    \longrightarrow
    \mathcal{L},
    \label{eq:forward_model}
\end{equation}
where $\Sigma$, $v_R$, and $v_\phi$ are the gas surface density and velocity fields at the final simulation time, $\hat{I}_{\rm raw}$ is the raw soft-binned map, and $\hat{I}$ is the processed map used in the loss calculation.
Figure~\ref{fig:gradient_propagation_chart} illustrates this forward calculation and the corresponding reverse-mode gradient propagation.
A more explicit description of the gradient path through the projected longitude, line-of-sight velocity, and emission weight of each grid cell is given in Section~\ref{subsec:objective_optimization}.

The gas fields are intermediate outputs of the hydrodynamical calculation and are not optimized as independent parameters.
Instead, the gradients pass through these fields and through the time integration to give the derivative of $\mathcal{L}$ with respect to $\Omega_{\rm b}$.
The viewing angle $\phi_{\rm b}$ and the gas response time $t_{\rm end}$ are treated as nuisance parameters in this paper and are explored through fixed choices or low-dimensional scans.

The model is intentionally simpler than a full three-dimensional, multi-phase, radiative-transfer calculation.
It is designed as a controlled forward model for testing the dynamical information contained in gas $\ell$--$v$ diagrams.

\subsection{Barred Milky Way potential}
\label{subsec:potential}

We assume that the gas moves in a prescribed, non-axisymmetric gravitational potential representing the barred Milky Way.
The potential is a composite Milky Way model consisting of a barred stellar component, an axisymmetric stellar disk, a nuclear stellar component, and a dark-matter halo.
For the barred stellar component, we use the stellar-dynamically constrained made-to-measure (M2M) model of \citet{Portail+2017}, adopting its analytic representation by \citet{Sormani+2022agama} as implemented in the {\tt Agama} framework \citep{Vasiliev2019}.
The remaining axisymmetric components are taken from the Milky Way model used in \citet{Baba2025b}, which is a slightly modified version of the model adopted by \citet{Hunter+2024}.
Thus, the shape and strength of the gravitational potential are fixed throughout this paper, and we use the gas response to test the pattern speed and viewing geometry of this fixed barred mass model.

The non-axisymmetric part of the potential rotates at a constant pattern speed $\Omega_{\rm b}$.
In the inertial frame, the total potential can be written as
\begin{equation}
    \Phi(R,\phi,t)
    =
    \Phi_{\rm axi}(R)
    +
    \Phi_{\rm bar}
    \left(R,\phi-\Omega_{\rm b} t\right),
    \label{eq:rotating_potential}
\end{equation}
where $(R,\phi)$ are Galactocentric polar coordinates in the Galactic plane, $\Phi_{\rm axi}$ denotes the axisymmetric part of the Milky Way potential, and $\Phi_{\rm bar}$ denotes the non-axisymmetric barred component.
In the frame corotating with the bar, the barred component is time independent.
In our coordinate convention, the Milky Way bar rotates in the negative azimuthal direction, so the fitted values of $\Omega_{\rm b}$ are negative.
When comparing with the literature, we often refer to the amplitude $|\Omega_{\rm b}|$ as the bar pattern speed.

In this work, $\Omega_{\rm b}$ is the primary parameter to be inferred.
Other structural parameters of the potential, including the bar mass distribution, disk component, nuclear stellar component, and dark-matter halo, are kept fixed.
The present-day bar orientation relative to the Sun--Galactic-center line is denoted by the viewing angle $\phi_{\rm b}$.
In several tests, $\phi_{\rm b}$ and the gas response time $t_{\rm end}$ are varied as nuisance parameters.

\subsection{Differentiable hydrodynamical solver}
\label{subsec:hydro}

We model the neutral gas as a non-self-gravitating, isothermal fluid in the Galactic plane and solve the two-dimensional hydrodynamical equations.
We adopt a constant effective sound speed $c_s=10\,{\rm km\,s^{-1}}$, which should be regarded as an effective turbulent velocity dispersion rather than the microscopic thermal sound speed of the gas.

We solve the gas dynamics in the frame corotating with the bar.
The gravitational potential $\Phi$ is the prescribed external Milky Way potential described in Section~\ref{subsec:potential}.
In this frame, the equations include the pressure force, the prescribed gravitational force, and the rotating-frame inertial forces, namely the Coriolis and centrifugal forces.
The conserved variables are the surface density, radial momentum, and angular momentum surface density in the bar's rotating frame.

Numerically, the gas equations are evolved on a polar grid with a differentiable finite-volume solver.
Numerical fluxes are computed with the Rusanov flux \citep{Rusanov1962}, a simple and robust approximate Riemann solver.
The source terms include pressure gradients, the prescribed gravitational force, and the rotating-frame inertial forces.
The time step is set by the Courant--Friedrichs--Lewy (CFL) condition, based on the local flow speed and the effective sound speed.
Reflecting boundaries are imposed in the radial direction, and periodic boundaries are imposed in the azimuthal direction.
All update operations are implemented as differentiable {\tt PyTorch} tensor operations, so that gradients can be propagated through the full time integration.

The initial gas state, $(\Sigma_0,\boldsymbol{v}_0)$, is smooth and axisymmetric.
Here $\Sigma_0$ is the initial gas surface density, and $\boldsymbol{v}_0$ is the initial gas velocity field.
In the fiducial setup, $\Sigma_0$ is specified by a Gaussian-ring radial profile with a small uniform background component.
The initial velocity field $\boldsymbol{v}_0$ has zero radial velocity and an azimuthal velocity set by circular rotation in the axisymmetric potential.
The same class of initial gas states is used in the mock tests and in the application to the observational data.
This initial condition is not intended to represent the present-day neutral-gas distribution.
Instead, it provides a simple starting point from which the bar-driven gas response is evolved. The comparison is therefore based on the developed non-axisymmetric response, rather than on the detailed initial radial profile.
During the evolution, the gas develops a non-axisymmetric response to the rotating barred potential, including shocks, offset dust-lane flows, and non-circular streaming motions.

The simulation is evolved until a gas response time $t_{\rm end}$, producing the state
\begin{equation}
    \left(
    \Sigma(R,\phi,t_{\rm end}),\;
    v_R(R,\phi,t_{\rm end}),\;
    v_\phi(R,\phi,t_{\rm end})
    \right),
\end{equation}
which is then passed to the observation operator described below.
Here $\Sigma$, $v_R$, and $v_\phi$ are the gas surface density, radial velocity, and azimuthal velocity at $t=t_{\rm end}$.
The parameter $t_{\rm end}$ does not represent the true age of the Galactic bar.
Rather, it controls the phase and degree of development of the gas response within our simplified forward model, and is therefore treated as a nuisance parameter in several tests.
In our code units, with lengths in kpc and velocities in ${\rm km\,s^{-1}}$, one time unit is approximately $0.98\,{\rm Gyr}$, which we round to $1\,{\rm Gyr}$.

\subsection{Differentiable $\ell$--$v$ projection}
\label{subsec:lv_projection}

As discussed in Section~\ref{sec:introduction}, the $\ell$--$v$ diagram retains the line-of-sight velocity structure of bar-driven gas flow.
We therefore compare the simulated gas response with the data directly in $\ell$--$v$ space.
Here we describe the differentiable projection operator used for this comparison.

To compare the simulated gas disk with observed CO data, we first convert the velocity from the bar's rotating frame to the inertial Galactocentric frame.
Here $\boldsymbol{u}_{\rm rot}\equiv (u_R^{\rm rot},u_\phi^{\rm rot})$ denotes the gas velocity measured in the frame corotating with the bar.
The inertial-frame velocity is given by $\boldsymbol{v}_{\rm inert} = \boldsymbol{u}_{\rm rot} + \boldsymbol{\Omega}_{\rm b}\times\boldsymbol{r}$,
or equivalently $v_R^{\rm inert}=u_R^{\rm rot}$ and $v_\phi^{\rm inert}=u_\phi^{\rm rot}+\Omega_{\rm b}R$.
We then project the gas state into the observer's $\ell$--$v$ space.
We place the Sun at $R_0=8.2\,{\rm kpc}$, adopt a local circular speed of $V_0=238\,{\rm km\,s^{-1}}$ \citep[][]{Bland-HawthornGerhard2016,GravityCollaboratio+2021}, and set the present-day bar viewing angle to $\phi_{\rm b}$.
For each polar grid cell, we rotate the gas map so that the bar major axis has the assumed angle $\phi_{\rm b}$ relative to the Sun--Galactic-center line, and then compute the Galactic longitude $\ell_n$ and the line-of-sight velocity $v_n$ relative to the local standard of rest.
Here $v$ in the $\ell$--$v$ map denotes the line-of-sight velocity.
The observer parameters $R_0$ and $V_0$ are held fixed, while $\phi_{\rm b}$ is treated as a viewing-geometry parameter.

A direct histogram of gas elements into $\ell$--$v$ bins is not differentiable with respect to the gas positions and velocities.
This is because a small change in $\ell_n$ or $v_n$ can move a cell abruptly from one pixel to another.
We therefore use a soft-binning operator in which each cell contributes to nearby $\ell$--$v$ pixels with smooth Gaussian kernel weights.
The predicted $\ell$--$v$ map is
\begin{equation}
    \hat{I}_{\rm raw}(\ell_i,v_j)
    =
    \sum_n
    W_n\,
    K_\ell(\ell_i-\ell_n)\,
    K_v(v_j-v_n).
\end{equation}
where the index $n$ labels polar grid cells, $\ell_n$ and $v_n$ are the longitude and line-of-sight velocity of cell $n$, $W_n$ is the emission weight, and $K_\ell$ and $K_v$ are Gaussian kernels in longitude and velocity.
They are defined as
\begin{equation}
    K_\ell(\Delta \ell)
    =
    \exp\left[
        -\frac{(\Delta \ell)^2}{2\sigma_\ell^2}
    \right]
    \label{eq:kl}
\end{equation}
and
\begin{equation}
    K_v(\Delta v)
    =
    \exp\left[
        -\frac{(\Delta v)^2}{2\sigma_v^2}
    \right].
    \label{eq:kv}
\end{equation}
Throughout the two-dimensional $\ell$--$v$ analyses in this paper, we use $\sigma_\ell=1.0^{\circ}$ and $\sigma_v=8.0\,{\rm km\,s^{-1}}$. The value $\sigma_\ell=1^\circ$ is chosen to emphasize coherent bar-driven structures rather than cloud-scale angular substructure. The velocity kernel accounts for unresolved turbulent motions and finite velocity-channel width.

For the two-dimensional $\ell$--$v$ comparisons, we set
\begin{equation}
    W_n \propto \Sigma_n A_n,
    \label{eq:weight_surface_density}
\end{equation}
where $A_n$ is the area of the grid cell.
This choice treats the projected surface density as an effective emission weight.
It is not intended to model the detailed radiative transfer or tracer-dependent emissivity of CO.

With the Gaussian kernels, each pixel value changes smoothly when $\ell_n$, $v_n$, or $W_n$ changes.
Thus, the projection step provides a differentiable connection between the hydrodynamical state and the raw predicted map $\hat{I}_{\rm raw}(\ell,v)$.

\subsection{Objective function and optimization}
\label{subsec:objective_optimization}

We compare the predicted $\ell$--$v$ map with a target map.
The target is either a self-consistency hydrodynamical mock, an independent hydrodynamical mock, or observed CO data.
These cases are analyzed in Sections~\ref{subsec:closed_loop}, \ref{subsec:sph_mock}, and \ref{sec:data_application}, respectively.

Before evaluating the loss, we place the model and target maps on the same $\ell$--$v$ grid and process them in the same way.
We write this processing step as
\begin{equation}
    \hat{I}
    =
    {\cal P}\!\left[\hat{I}_{\rm raw}\right],
    \label{eq:model_processing}
\end{equation}
where $\hat{I}_{\rm raw}$ is the raw soft-binned model map defined in Section~\ref{subsec:lv_projection}, and $\hat{I}$ is the processed model map used in the loss calculation.
In this paper, the operator ${\cal P}$ denotes the processing applied before the loss is evaluated. 
It includes logarithmic compression and robust percentile-based normalization of the model and target map intensities.
For the observed CO application, we additionally add a small intensity floor before logarithmic compression.
The target map is processed in the same way and is denoted by $I^{\rm target}$.

Thus, the loss is evaluated on processed and normalized morphology maps, not on the raw brightness temperature or raw projected surface density.
This choice follows the argument of \citet{SormaniMagorrian2015} that broad $\ell$--$v$ features trace the large-scale gas dynamics more robustly than detailed brightness distributions.
Accordingly, $\mathcal{L}$ should be interpreted as a morphology-based objective function, not as a formal pixel-by-pixel likelihood.
Its absolute scale depends on the processing operator ${\cal P}$, the fitting mask, and model mismatch, so we use it only for relative comparisons within the same setup.

This motivates a loss function that is insensitive to the overall intensity scale but sensitive to the alignment of large-scale structures in the processed map.
We therefore adopt a cosine-distance loss.
The mask enters this definition as a restriction of the pixel space: equivalently, we use the masked inner product
$\langle A,B\rangle_M=\sum_{i,j}M_{ij}A_{ij}B_{ij}$ and the corresponding norm $\|A\|_M=\langle A,A\rangle_M^{1/2}$.
The loss is then one minus the cosine similarity between the processed model and target maps in this masked pixel space,
\begin{equation}
    \mathcal{L}
    =
    1
    -
    \frac{
    \sum_{i,j}
    M_{ij}
    \hat{I}_{ij}
    I^{\rm target}_{ij}
    }{
    \left[
    \sum_{i,j}
    M_{ij}
    \hat{I}_{ij}^{\,2}
    \right]^{1/2}
    \left[
    \sum_{i,j}
    M_{ij}
    \left(I^{\rm target}_{ij}\right)^2
    \right]^{1/2}
    } .
    \label{eq:loss_cos}
\end{equation}
Here $i$ and $j$ label longitude and velocity pixels, and $M_{ij}$ is the fitting mask, with $M_{ij}=1$ for pixels included in the comparison and $M_{ij}=0$ otherwise.
Thus, $\mathcal{L}$ measures the angular mismatch between the processed model and target maps after restricting both maps to the fitting region.

This loss is minimized when the processed model and target maps have the same direction in the masked pixel space.
It mainly measures whether the two maps have similar large-scale $\ell$--$v$ morphology within the fitting mask.
This choice avoids introducing an additional weighting hyperparameter between different loss terms and is appropriate for the present proof-of-concept study, which aims to test whether coherent gas-dynamical structures in $\ell$--$v$ space can constrain the bar pattern speed.
The mask allows us to emphasize informative regions, such as the terminal-velocity region, and to exclude regions dominated by local gas emission, low survey sensitivity, or small-scale structures that are not included in the model.

The differentiable projection in Section~\ref{subsec:lv_projection} connects the objective function to the dynamical parameters through the computational graph.
For the pattern-speed parameter, the gradient is propagated schematically as
\begin{equation}
    \frac{\partial \mathcal{L}}{\partial \hat{I}}
    \longrightarrow
    \frac{\partial \mathcal{L}}{\partial \hat{I}_{\rm raw}}
    \longrightarrow
    \frac{\partial \mathcal{L}}{\partial (\ell_n, v_n, W_n)}
    \longrightarrow
    \frac{\partial \mathcal{L}}{\partial (\Sigma, v_R, v_\phi)}
    \longrightarrow
    \frac{\partial \mathcal{L}}{\partial \Omega_{\rm b}} .
    \label{eq:gradient_flow_objective}
\end{equation}
The first step passes through the processing operator ${\cal P}$.
Here $\ell_n$, $v_n$, and $W_n$ are the projected longitude, line-of-sight velocity, and emission weight used to construct $\hat{I}_{\rm raw}$.
The Gaussian soft-binning operator makes this projection step differentiable by making the raw predicted map a smooth function of these quantities.
Automatic differentiation then propagates the gradient through the hydrodynamical time integration to obtain $\partial \mathcal{L}/\partial \Omega_{\rm b}$.
Thus, the data mismatch can provide gradients for the bar parameters.

Because the full forward model is differentiable, we use $\partial \mathcal{L}/\partial \Omega_{\rm b}$ to update $\Omega_{\rm b}$ with the Adam optimizer \citep{KingmaBa2015}.
We start the optimization from several initial values of $\Omega_{\rm b}$ to test whether the solution converges to the same low-loss region.
As a check, we also compute explicit scans of $\mathcal{L}$ over $\Omega_{\rm b}$.
These scans help us identify secondary minima caused by transient gas response, viewing geometry, or mask choices.

In some experiments, we also compute grids over $(\Omega_{\rm b},t_{\rm end})$ or $(\Omega_{\rm b},\phi_{\rm b})$.
These low-dimensional scans do not require differentiability, but they are useful for visualizing parameter degeneracies and interpreting the gradient-based results.
The advantage of the differentiable approach is that it can be extended to higher-dimensional parameter spaces, where exhaustive grid searches become impractical.

\section{Mock validation}
\label{sec:mock_validation}

Before applying the differentiable forward model to observed data, we validate it with two types of mock data.
The first is a self-consistency hydrodynamical mock, in which the target $\ell$--$v$ map is generated by the same differentiable solver used in the fitting model.
This test contains no model mismatch and is designed to verify the computational graph, the gradient calculation, and the optimizer.

The second is an independent hydrodynamical mock based on simulations performed with the $N$-body/SPH code \texttt{ASURA} \citep{Saitoh+2008, SaitohMakino2013} and used in \citet{Baba2025b, Baba2026a}.
Although these simulations adopt the same barred Milky Way potential as the fitting model, their gas dynamics are much more realistic.
They include gas self-gravity, radiative cooling, star formation, and stellar feedback.
They also differ from the fitting model in numerical method, resolution, thermodynamics, and small-scale gas structure.
This test asks whether the differentiable forward model can recover the large-scale bar-driven kinematic signal from a realistic gas distribution that was not generated by the fitting model.

Together, these tests examine whether the framework recovers the correct pattern speed under ideal conditions and still captures the large-scale bar-driven kinematic signal under model mismatch.

\subsection{Self-consistency mock validation}
\label{subsec:closed_loop}

We first perform a self-consistency test using the differentiable hydrodynamical solver itself.
Here, ``self-consistency'' means that the target $\ell$--$v$ map and the fitting model are generated with the same differentiable forward solver, so that there is no model mismatch.
The target $\ell$--$v$ map is generated with a fixed true pattern-speed parameter, $\Omega_{\rm b}^{\rm true}$.
Starting from a different initial value, $\Omega_{\rm b}^{\rm init}$, we optimize $\Omega_{\rm b}$ and test whether it converges toward $\Omega_{\rm b}^{\rm true}$.
This test checks the differentiable implementation, the projection operator, and the optimization procedure under ideal conditions.

The polar grid and barred potential are set up as described in Section~\ref{sec:methods}.
For this test, we use a $128 \times 128$ grid covering $0.5 \leq R \leq 10\,{\rm kpc}$ and $0 \leq \phi \leq 2\pi$, and fix the bar viewing angle to $\phi_{\rm b}=25^{\circ}$.
The target and fitting runs are evolved from the same axisymmetric initial disk to the same final response time $t_{\rm end}$, which is fixed and not optimized in the self-consistency tests.
The model and target maps are then projected onto the same $\ell$--$v$ grid using the soft-binning prescription defined in Section~\ref{subsec:lv_projection}, processed in the same way as defined in Section~\ref{subsec:objective_optimization}, and compared with the loss function.

\begin{figure}
\begin{center}
\includegraphics[width=0.48\textwidth]{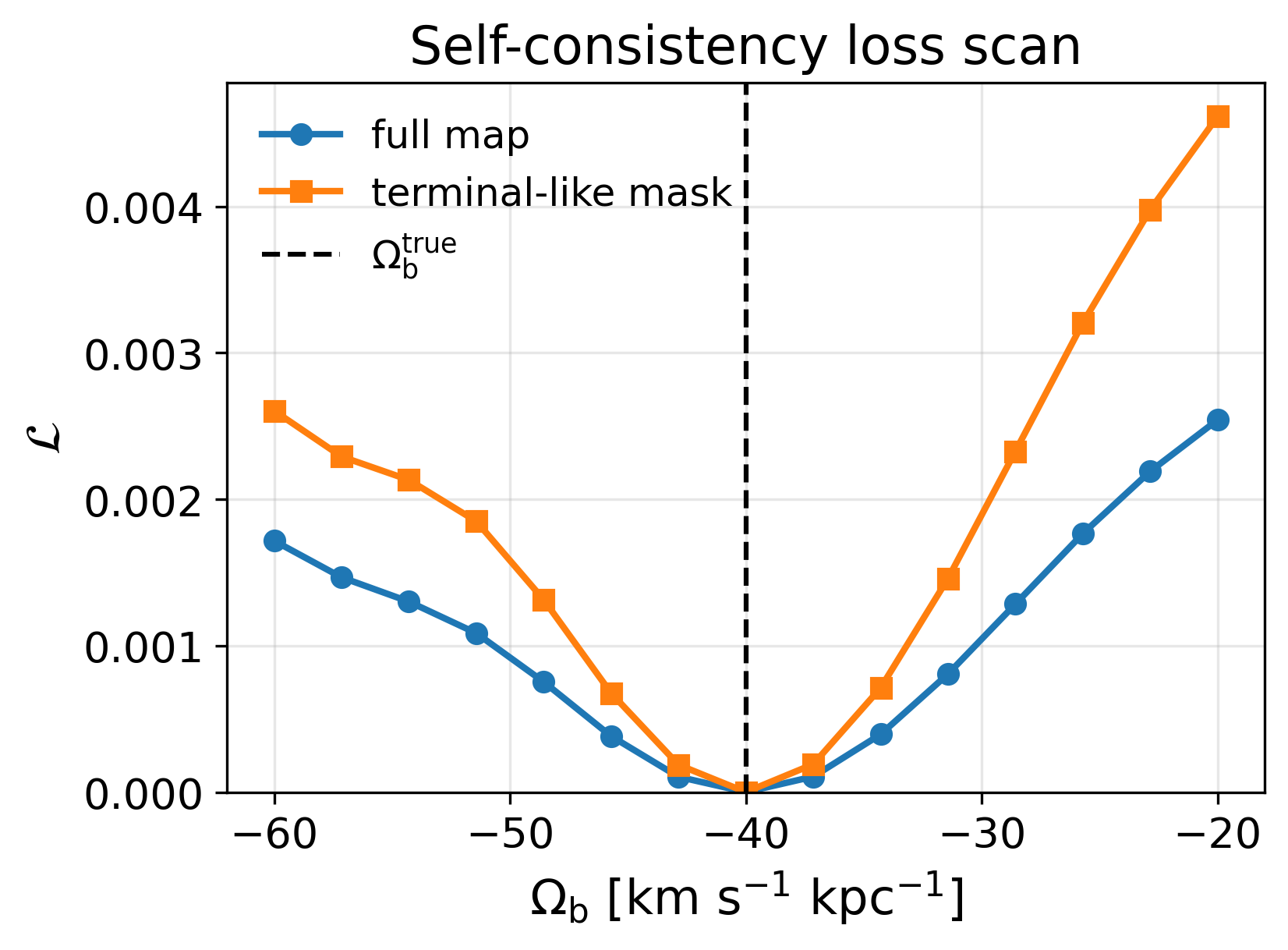}
\end{center}
\caption{
Self-consistency loss scan for a hydrodynamical target generated with $\Omega_{\rm b}^{\rm true}=-40\,{\rm km\,s^{-1}\,kpc^{-1}}$.
In this test, the target map and the fitting models are generated with the same differentiable forward solver, so there is no model mismatch.
The cosine-distance loss $\mathcal{L}$ is shown as a function of the fitted pattern-speed parameter.
Circles show the full-map comparison, and squares show the terminal-like mask.
The vertical dashed line marks the input value.
Both masks give a clear minimum at the true pattern speed, with a minimum loss close to zero.
\textbf{Alt text:} Loss curves for two fitting masks in a self-consistency mock test. Both curves reach a clear minimum at the true bar pattern speed, $-40\,{\rm km\,s^{-1}\,kpc^{-1}}$.
}
\label{fig:hydro_loss_scan}
\end{figure}

Figure~\ref{fig:hydro_loss_scan} shows a representative scan of $\mathcal{L}$ for a target with $\Omega_{\rm b}^{\rm true}=-40\,{\rm km\,s^{-1}\,kpc^{-1}}$.
For both the full-map and terminal-like masks, the cosine-distance loss has a clear minimum at the true value and increases on both sides over the explored range from $\Omega_{\rm b}=-60$ to $-20\,{\rm km\,s^{-1}\,kpc^{-1}}$.
The minimum loss is very close to zero, as expected for a self-consistency test in which the target map can be reproduced by the fitting model.
The two masks give similar loss curves in this no-mismatch case.
This indicates that the differentiable solver and the soft-binning operator recover the input pattern speed under ideal conditions.
The terminal-like mask is not essential for this idealized test, but it is useful as a controlled reference for the more realistic mock and observational applications below.

\begin{figure*}
\begin{center}
\includegraphics[width=0.95\textwidth]{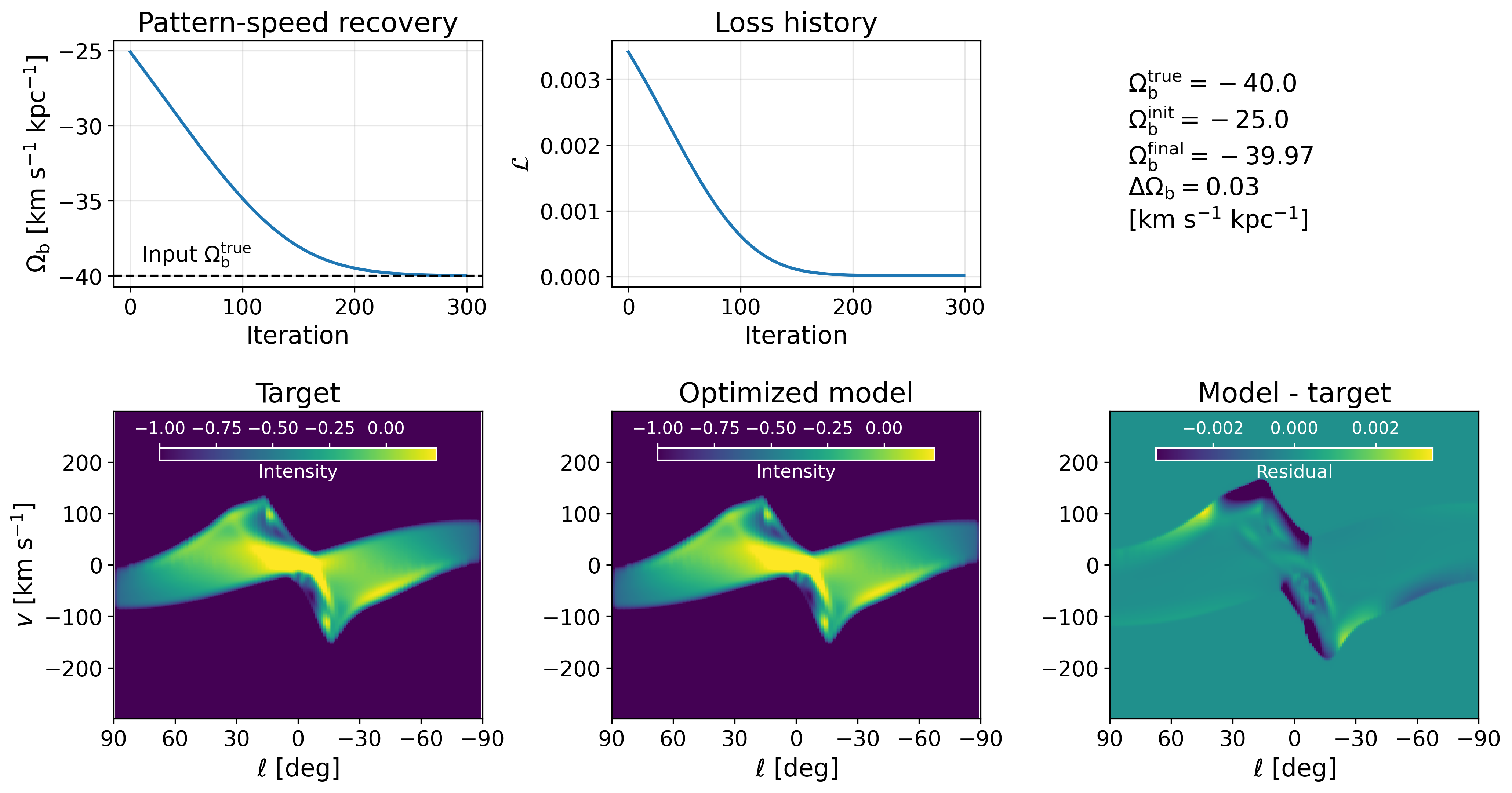}
\end{center}
\caption{
Representative optimization run in the self-consistency mock test.
The target map is generated with $\Omega_{\rm b}^{\rm true}=-40\,{\rm km\,s^{-1}\,kpc^{-1}}$, while the optimization starts from $\Omega_{\rm b}^{\rm init}=-25\,{\rm km\,s^{-1}\,kpc^{-1}}$.
The upper panels show the evolution of $\Omega_{\rm b}$ and the cosine-distance loss $\mathcal{L}$ during the Adam optimization, together with the final recovered value.
The horizontal dashed line in the pattern-speed panel marks the input value, $\Omega_{\rm b}^{\rm true}$.
The lower panels compare the target $\ell$--$v$ map, the optimized model map, and the residual.
The intensity scale denotes the processed, dimensionless map intensity used in the loss calculation, not a physical brightness temperature.
The residual is defined as the difference between the processed model and target maps, and is therefore dimensionless.
The recovered pattern speed is $\Omega_{\rm b}^{\rm final}=-39.97\,{\rm km\,s^{-1}\,kpc^{-1}}$, with a bias of $\Delta\Omega_{\rm b}=0.03\,{\rm km\,s^{-1}\,kpc^{-1}}$.
\textbf{Alt text:} Optimization history and final map comparison for a self-consistency mock test. The pattern speed converges from $-25$ to nearly $-40\,{\rm km\,s^{-1}\,kpc^{-1}}$, the loss decreases, and the optimized longitude--velocity map closely reproduces the target map with small residuals.
}
\label{fig:hydro_single}
\end{figure*}

Figure~\ref{fig:hydro_single} shows a representative optimization run with $\Omega_{\rm b}^{\rm true}=-40\,{\rm km\,s^{-1}\,kpc^{-1}}$ and $\Omega_{\rm b}^{\rm init}=-25\,{\rm km\,s^{-1}\,kpc^{-1}}$.
The cosine-distance loss decreases rapidly during the first $\sim 100$ iterations and becomes nearly zero by $\sim 200$ iterations.
At the same time, the fitted pattern speed converges smoothly toward the true value.
The recovered value is $\Omega_{\rm b}^{\rm final}=-39.97\,{\rm km\,s^{-1}\,kpc^{-1}}$, corresponding to a bias of $\Delta\Omega_{\rm b}=0.03\,{\rm km\,s^{-1}\,kpc^{-1}}$.
The optimized $\ell$--$v$ map closely reproduces the target map.
The residuals are small compared with the dynamic range of the processed maps, with the remaining coherent residuals mainly confined to sharp emission features.
This behavior confirms that the cosine-distance objective provides useful gradients for recovering the pattern speed when the model class is correct.

We also repeated the self-consistency test for several true pattern speeds and initial guesses.
Specifically, we used true pattern speeds of $\Omega_{\rm b}^{\rm true}=-35$, $-40$, and $-45\,{\rm km\,s^{-1}\,kpc^{-1}}$,
and initial guesses of $\Omega_{\rm b}^{\rm init}=-20$, $-25$, $-30$, and $-50\,{\rm km\,s^{-1}\,kpc^{-1}}$.
Table~\ref{tab:closed_loop_recovery} summarizes the recovered values for the terminal-like mask.
The optimizer recovers the input pattern speed accurately over this range of tests.
For $\Omega_{\rm b}^{\rm true}=-35$ and $-40\,{\rm km\,s^{-1}\,kpc^{-1}}$, the absolute bias is at most $0.07\,{\rm km\,s^{-1}\,kpc^{-1}}$.
For the fastest case, $\Omega_{\rm b}^{\rm true}=-45\,{\rm km\,s^{-1}\,kpc^{-1}}$, the recovery remains accurate but shows a slightly larger bias when the optimization starts from much slower initial values, reaching $0.76\,{\rm km\,s^{-1}\,kpc^{-1}}$ for $\Omega_{\rm b}^{\rm init}=-20\,{\rm km\,s^{-1}\,kpc^{-1}}$.

These results show that the differentiable solver, soft-binning projection, cosine-distance loss, and Adam optimization recover the input pattern speed under ideal conditions with no model mismatch.
They also provide the baseline for the more challenging tests below, where the target data are generated by an independent SPH simulation or taken from observed CO surveys.

\begin{table*}
\caption{Recovered bar pattern speed in the self-consistency test using the terminal-like mask. Columns give the initial value $\Omega_{\rm b}^{\rm init}$, and rows give the true value $\Omega_{\rm b}^{\rm true}$. Each entry lists the recovered $\Omega_{\rm b}$, with the difference from the true value in parentheses. All values are in $\mathrm{km\,s^{-1}\,kpc^{-1}}$.}
\label{tab:closed_loop_recovery}
\begin{center}
\begin{tabular}{lrrrr}
\hline
$\Omega_{\rm b}^{\rm true}$ $\backslash$ $\Omega_{\rm b}^{\rm init}$
& -50.0 & -30.0 & -25.0 & -20.0 \\
\hline
-45.0 & -45.00 (+0.00) & -44.88 (+0.12) & -44.55 (+0.45) & -44.24 (+0.76) \\
-40.0 & -40.00 (+0.00) & -40.00 (+0.00) & -39.99 (+0.01) & -39.93 (+0.07) \\
-35.0 & -35.02 (-0.02) & -35.00 (+0.00) & -35.00 (+0.00) & -35.00 (-0.00) \\
\hline
\end{tabular}
\end{center}
\end{table*}

\subsection{Independent mock validation}
\label{subsec:sph_mock}

We next test the method against an independent hydrodynamical mock introduced above.
This test keeps the same barred potential family but replaces the ideal self-consistency target with a more realistic gas distribution generated by a different numerical scheme and more complete interstellar-medium physics \citep{Baba2025b, Baba2026a}.
Specifically, the target is generated from an SPH simulation whose adopted bar pattern speed is $\Omega_{\rm b}^{\rm mock}=-37.5\,{\rm km\,s^{-1}\,kpc^{-1}}$, which we use as the reference value.
We construct the target $\ell$--$v$ map from gas particles representative of the neutral component, selected by $T<10^4\,{\rm K}$, $n_{\rm H}>10^{-2}\,{\rm cm^{-3}}$, and $|z|<0.1\,{\rm kpc}$.
These particles are projected into $\ell$--$v$ space using the same Gaussian soft-binning prescription as in Section~\ref{subsec:lv_projection}, except that the emission weight $W_n$ is replaced by the particle mass.
As in the self-consistency test, the target and model maps are processed in the same way using the operator ${\cal P}$ defined in Section~\ref{subsec:objective_optimization} before evaluating the loss.

The fitting model is the differentiable hydrodynamic solver described in Section~\ref{sec:methods}.
For the fiducial independent-mock comparison, we use a $128\times128$ polar grid covering radii from $R=0.5$ to $10\,{\rm kpc}$.
Unless otherwise stated, we adopt $\phi_{\rm b}=25^{\circ}$ and set the gas response time to $t_{\rm end}=0.30$, corresponding to $\approx300\,{\rm Myr}$.
We adopt this response time because the main bar-driven shocks and elongated $\ell$--$v$ structures are already established and evolve more slowly than during the initial transient phase, although the response is not strictly time independent.
We do not assume that this value is uniquely preferred; the dependence on $t_{\rm end}$ is examined explicitly in Section~\ref{subsec:mock_degeneracies}.

\begin{figure}
\begin{center}
\includegraphics[width=0.48\textwidth]{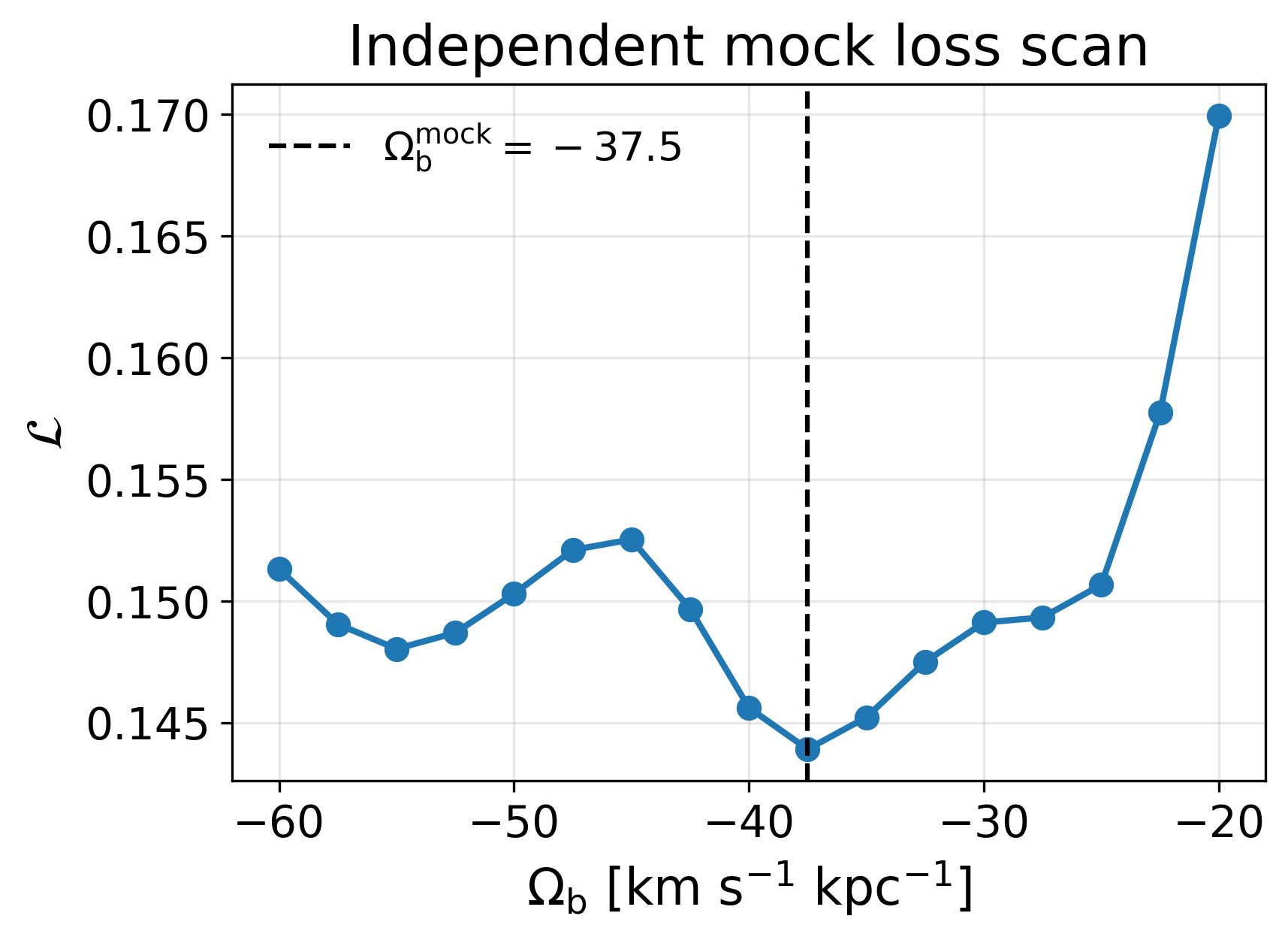}
\end{center}
\caption{
Loss scan for the independent hydrodynamical mock.
The loss $\mathcal{L}$ is shown as a function of the fitted pattern-speed parameter for the fiducial differentiable model with radii from $R=0.5$ to $10\,{\rm kpc}$, $\phi_{\rm b}=25^\circ$, and $t_{\rm end}=0.30$.
This response time is chosen so that the differentiable gas model is in a quasi-steady barred-flow state, rather than in the early transient phase.
The vertical dashed line marks the input mock value, $\Omega_{\rm b}^{\rm mock}=-37.5\,{\rm km\,s^{-1}\,kpc^{-1}}$.
The minimum lies close to the input mock value, but it is broader and more asymmetric than in the self-consistency test because the independent target and the differentiable fitting model differ in gas physics, numerical method, and small-scale structure.
\textbf{Alt text:} Loss scan for the independent hydrodynamical mock. The loss reaches its lowest value close to the input mock pattern speed of $-37.5\,{\rm km\,s^{-1}\,kpc^{-1}}$, but the minimum is broad and asymmetric because the target simulation and the differentiable fitting model are not identical.
}
\label{fig:sph_loss_scan}
\end{figure}

Figure~\ref{fig:sph_loss_scan} shows a scan of $\mathcal{L}$ as a function of $\Omega_{\rm b}$ for the independent target in this fiducial setup.
Despite the mismatch between the independent hydrodynamical target and the differentiable grid model, the loss has its minimum very close to the input value, $\Omega_{\rm b}^{\rm mock}=-37.5\,{\rm km\,s^{-1}\,kpc^{-1}}$.
The minimum is much broader than in the self-consistency test, reflecting the numerical and physical mismatch between the two models.
The loss increases steeply toward substantially slower bars, especially for $|\Omega_{\rm b}|\lesssim25$--$30\,{\rm km\,s^{-1}\,kpc^{-1}}$.
On the fast-bar side, the loss varies more gradually and shows secondary structure, indicating that some fast models can still reproduce part of the large-scale morphology.
Thus, the independent mock identifies the correct input pattern speed as the lowest-loss value in this one-dimensional scan, but the broad and asymmetric loss curve shows that model mismatch prevents a sharply localized constraint.
The result demonstrates that the large-scale $\ell$--$v$ structure of the independent hydrodynamical mock retains useful information on $\Omega_{\rm b}$ even under model mismatch.

\begin{figure*}
\begin{center}
\includegraphics[width=0.95\textwidth]{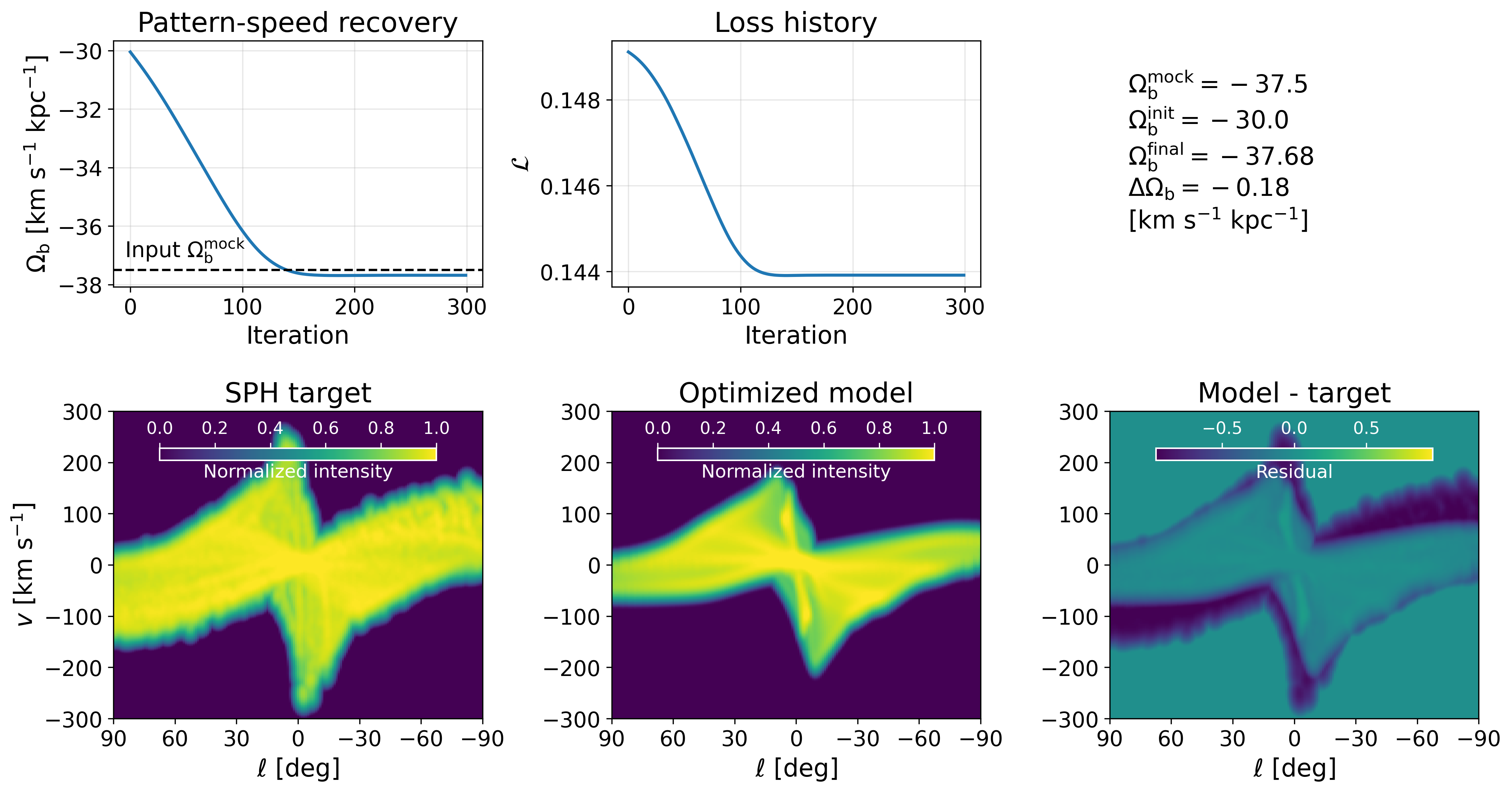}
\end{center}
\caption{
Representative optimization run for the independent hydrodynamical mock.
The independent target has an input pattern speed $\Omega_{\rm b}^{\rm mock}=-37.5\,{\rm km\,s^{-1}\,kpc^{-1}}$,
while the optimization starts from $\Omega_{\rm b}^{\rm init}=-30\,{\rm km\,s^{-1}\,kpc^{-1}}$.
The gas response time is fixed to $t_{\rm end}=0.30$.
The upper panels show the evolution of $\Omega_{\rm b}$ and $\mathcal{L}$ during the Adam optimization, together with the final recovered value.
The horizontal dashed line in the pattern-speed panel marks the input SPH mock value, $\Omega_{\rm b}^{\rm mock}$.
The fitted pattern speed converges to $\Omega_{\rm b}^{\rm final}=-37.68\,{\rm km\,s^{-1}\,kpc^{-1}}$,
close to the input mock value and within the broad low-loss region found in Figure~\ref{fig:sph_loss_scan}.
The lower panels compare the SPH target $\ell$--$v$ map, the optimized differentiable model, and the residual.
The intensity scale denotes the processed, dimensionless map intensity used in the loss calculation.
The residual is the difference between the processed model and target maps.
Coherent residuals remain because the independent target contains small-scale structure and physical processes that are not included in the simplified differentiable model.
\textbf{Alt text:} Optimization history and final map comparison for an independent hydrodynamical mock. The fitted pattern speed converges close to the input mock value, but the residual map still shows coherent differences between the SPH target and the simplified differentiable model.
}
\label{fig:sph_single}
\end{figure*}

Figure~\ref{fig:sph_single} shows a representative gradient-based optimization run initialized at $\Omega_{\rm b}^{\rm init}=-30\,{\rm km\,s^{-1}\,kpc^{-1}}$.
This run uses the same quasi-steady response time as the fiducial independent-mock scan, $t_{\rm end}=0.30$.
The cosine-distance loss decreases during the first $\sim 100$ iterations and then becomes nearly flat.
At the same time, the fitted pattern speed moves from the initial value toward the input mock value and the low-loss region identified in Figure~\ref{fig:sph_loss_scan}.
The recovered value is $\Omega_{\rm b}^{\rm final}=-37.68\,{\rm km\,s^{-1}\,kpc^{-1}}$, within $0.18\,{\rm km\,s^{-1}\,kpc^{-1}}$ of the input mock value.
Thus, the optimization recovers the input pattern speed of the independent mock in this representative setup.

The optimized model reproduces the main large-scale $\ell$--$v$ morphology of the independent target, including the broad elongated emission pattern and the main high-velocity structures.
However, it does not reproduce the clumpy small-scale structure, the broader emission envelope, and the detailed high-velocity features present in the SPH target.
The residuals therefore show coherent structure, reflecting the physical and numerical mismatch between the independent hydrodynamical simulation and the simplified differentiable model.
This behavior is expected for the independent mock and contrasts with the nearly exact recovery in the self-consistency test.

Thus, the independent-mock validation should not be read as a pixel-level reproduction test.
The independent target contains clumps, feedback-driven structures, multiphase gas, and numerical small-scale features that are absent from the simplified differentiable model.
These differences produce coherent residuals and broaden the loss minimum.
The key test is instead whether the model can identify the correct large-scale bar-driven response in $\ell$--$v$ space under physical and numerical model mismatch.
The recovered pattern speed shows that the dominant kinematic signal of the rotating bar is still captured, even though the detailed gas morphology is not.
This distinction is important for the observational application, where tracer-dependent small-scale structures are also expected to remain imperfectly modeled.

\begin{table*}
\caption{
Recovered bar pattern speed for the independent hydrodynamical mock test at $t_{\rm end}=0.30$.
Columns give the initial value $\Omega_{\rm b}^{\rm init}$, and rows give the fitting mask.
Each entry lists the recovered $\Omega_{\rm b}^{\rm final}$, with
$\Delta\Omega_{\rm b}=\Omega_{\rm b}^{\rm final}-\Omega_{\rm b}^{\rm mock}$
in parentheses.
The input mock value is
$\Omega_{\rm b}^{\rm mock}=-37.5\,{\rm km\,s^{-1}\,kpc^{-1}}$.
All values are in $\mathrm{km\,s^{-1}\,kpc^{-1}}$.
}
\label{tab:sph_mock_recovery}
\begin{center}
\begin{tabular}{lrrrrr}
\hline
Mask $\backslash$ $\Omega_{\rm b}^{\rm init}$
& -50.0 & -45.0 & -40.0 & -30.0 & -25.0 \\
\hline
Full map
& -40.04 (-2.54)
& -40.00 (-2.50)
& -40.01 (-2.51)
& -40.00 (-2.50)
& -25.65 (+11.85) \\
Terminal-like
& -55.04 (-17.54)
& -37.70 (-0.20)
& -37.65 (-0.15)
& -37.70 (-0.20)
& -26.08 (+11.42) \\
\hline
\end{tabular}
\end{center}
\end{table*}

\subsection{Degeneracies and robustness checks}
\label{subsec:mock_degeneracies}

The independent hydrodynamical mock also allows us to examine the main degeneracies that affect the interpretation of real H\,{\sc i} and CO data.
The $\ell$--$v$ structure depends not only on $\Omega_{\rm b}$, but also on the bar viewing angle $\phi_{\rm b}$, the gas response time $t_{\rm end}$, and the radial domain of the gas calculation.
We therefore compute additional loss scans in which these quantities are varied.

\begin{figure*}
\begin{center}
\includegraphics[width=0.95\textwidth]{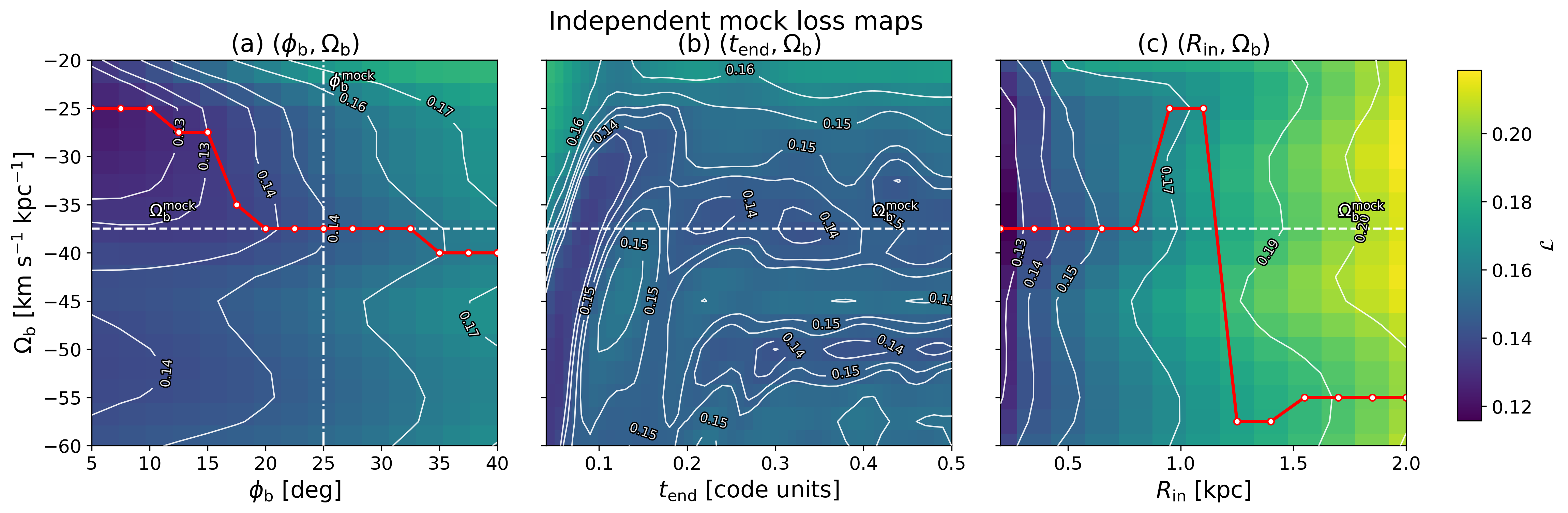}
\end{center}
\caption{
Degeneracy tests for the independent hydrodynamical mock.
Panel (a) shows the loss map in the $(\phi_{\rm b},\Omega_{\rm b})$ plane at fixed $t_{\rm end}=0.30$ and $R_{\rm in}=0.5\,{\rm kpc}$.
Panel (b) shows the loss map in the $(t_{\rm end},\Omega_{\rm b})$ plane at fixed $\phi_{\rm b}=25^{\circ}$ and $R_{\rm in}=0.5\,{\rm kpc}$.
Panel (c) shows the loss map in the $(R_{\rm in},\Omega_{\rm b})$ plane at fixed $t_{\rm end}=0.30$ and $\phi_{\rm b}=25^{\circ}$.
The background images show the loss $\mathcal{L}$, with contour lines indicating selected loss levels.
The horizontal dashed lines mark the input mock pattern speed,
$\Omega_{\rm b}^{\rm mock}=-37.5\,{\rm km\,s^{-1}\,kpc^{-1}}$.
The vertical dot-dashed line in panel (a) marks the input mock viewing angle,
$\phi_{\rm b}^{\rm mock}=25^{\circ}$.
The red circles connected by solid lines in panels (a) and (c) show the minimum-loss value of $\Omega_{\rm b}$ at each fixed $\phi_{\rm b}$ or $R_{\rm in}$.
\textbf{Alt text:} Three two-dimensional loss maps for the independent hydrodynamical mock. Panel (a) shows the dependence on bar viewing angle and pattern speed, panel (b) on gas response time and pattern speed, and panel (c) on the inner gas radius and pattern speed. The input mock pattern speed is marked in each panel, and the low-loss regions are broad, showing degeneracies between the pattern speed and the other model choices.
}
\label{fig:sph_degeneracy_loss_maps}
\end{figure*}

First, we vary the bar viewing angle over $\phi_{\rm b}=5^{\circ}$--$40^{\circ}$ at fixed $t_{\rm end}=0.30$ and $R_{\rm in}=0.5\,{\rm kpc}$.
Figure~\ref{fig:sph_degeneracy_loss_maps}a shows the two-dimensional loss map in the $(\phi_{\rm b},\Omega_{\rm b})$ plane.
The low-loss region forms a broad ridge that passes near the input mock pattern speed and viewing angle, $\Omega_{\rm b}^{\rm mock}=-37.5\,{\rm km\,s^{-1}\,kpc^{-1}}$ and $\phi_{\rm b}^{\rm mock}=25^{\circ}$.
The minimum-loss track shifts systematically with viewing angle: it favors smaller $|\Omega_{\rm b}|$ at small $\phi_{\rm b}$, passes close to the mock value at intermediate viewing angles, and moves toward larger $|\Omega_{\rm b}|$ at larger $\phi_{\rm b}$.
This shift arises because $\phi_{\rm b}$ changes the projection geometry of the same barred gas flow.
For a fixed gas response stage, changing $\phi_{\rm b}$ moves the bar-driven streaming motions, shocks, and high-velocity ridges to different longitudes and line-of-sight velocities.
Changing $\Omega_{\rm b}$ also changes these structures, because it changes the phase and strength of the gas response to the rotating bar.
Thus, different combinations of $\Omega_{\rm b}$ and $\phi_{\rm b}$ can place similar large-scale kinematic features in similar regions of the $\ell$--$v$ diagram.
This shows that $\Omega_{\rm b}$ and $\phi_{\rm b}$ are partially degenerate in the $\ell$--$v$ comparison, even though the independent mock still retains a broad low-loss region near the correct pattern speed and viewing angle.

Second, we vary the gas response time over $t_{\rm end}=0.04$--$0.50$ code units, corresponding to approximately $40$--$500\,{\rm Myr}$.
Figure~\ref{fig:sph_degeneracy_loss_maps}b shows the two-dimensional loss map in the $(t_{\rm end},\Omega_{\rm b})$ plane at fixed $\phi_{\rm b}=25^{\circ}$.
The loss map is structured and does not show a single narrow global valley.
Around $t_{\rm end}\simeq0.2$--$0.4$, several low-loss regions occur near the input mock value, $\Omega_{\rm b}^{\rm mock}=-37.5\,{\rm km\,s^{-1}\,kpc^{-1}}$.
Additional low-loss branches also appear at larger $|\Omega_{\rm b}|$ for some response times.
This behavior shows that the gas response time can introduce multiple low-loss branches, in addition to a local degeneracy with $\Omega_{\rm b}$.
Physically, this means that the pattern speed and the response stage of the gas can partly compensate for each other: different pairs of $(\Omega_{\rm b},t_{\rm end})$ can produce similar large-scale $\ell$--$v$ morphology, even though they correspond to different gas-flow histories.
Therefore, the minimum in the $(t_{\rm end},\Omega_{\rm b})$ loss map should not be interpreted as a unique recovery of $\Omega_{\rm b}$ unless the allowed range of $t_{\rm end}$ is physically constrained.

Finally, we test the sensitivity to the radial domain of the differentiable gas calculation.
Here $R_{\rm in}$ denotes the inner boundary of the polar grid.
The gas response in the inner few kiloparsecs is strongly affected by shocks and orbit crowding near the bar.
We therefore compute a two-dimensional loss map in the $(R_{\rm in},\Omega_{\rm b})$ plane, varying $R_{\rm in}$ over $0.2$--$2.0\,{\rm kpc}$ while keeping $t_{\rm end}=0.30$, $\phi_{\rm b}=25^{\circ}$, and the outer boundary fixed.
Figure~\ref{fig:sph_degeneracy_loss_maps}c shows that the loss depends strongly on the adopted inner boundary.
For small inner boundaries, the minimum-loss track remains close to the input mock value.
For intermediate values around $R_{\rm in}\simeq1.0$--$1.1\,{\rm kpc}$, the preferred pattern speed shifts toward smaller $|\Omega_{\rm b}|$.
For larger inner boundaries, the minimum-loss track moves to a much faster branch.
This behavior indicates that the inner radial domain contains important information for constraining the bar-driven $\ell$--$v$ morphology.
Excluding part of the inner Galaxy changes not only the amount of gas included in the calculation, but also the shocks, orbit-crowding features, and high-velocity structures that carry much of the pattern-speed information.
Thus, the inferred pattern speed is sensitive not only to $\phi_{\rm b}$ and $t_{\rm end}$, but also to the adopted radial domain of the gas calculation.

In summary, the mock tests show that the differentiable forward model passes two basic requirements needed before applying it to real data.
First, it recovers the true pattern speed in a self-consistency setting where there is no model mismatch.
Second, it identifies a broad low-loss region around the correct pattern speed even when the target is generated by an independent hydrodynamical simulation with different gas physics and a different numerical scheme.
The robustness checks also show that the inferred pattern speed is partially degenerate with $\phi_{\rm b}$ and $t_{\rm end}$, that the loss landscape can contain multiple low-loss branches, and that the result is sensitive to the adopted radial domain of the gas calculation.
These results motivate the cautious interpretation of the observed-data application below.

\section{Application to observational data}
\label{sec:data_application}

Having validated the differentiable forward model with self-consistency hydrodynamical mocks and an independent hydrodynamical mock in Section~\ref{sec:mock_validation}, we now apply the same framework to observed Galactic gas data.
We use the composite CO survey of \citet{Dame+2001} and focus on the molecular gas $\ell$--$v$ structure of the inner Milky Way.
This application should be regarded as a first consistency test on real data, not as a final precision measurement of $\Omega_{\rm b}$.
The Dame et al. CO cube, from which we construct the observed latitude-integrated CO $\ell$--$v$ map, contains emission from local gas, the Central Molecular Zone (CMZ), spiral arms, optical-depth effects, tracer-dependent emissivity variations, and survey-coverage effects that are not represented in the simplified forward model.
We therefore focus on large-scale bar-driven kinematic structure in $\ell$--$v$ space and on broad low-loss regions in $(\Omega_{\rm b},\phi_{\rm b})$, rather than on a single best-fitting value.

\subsection{Data and setup}
\label{subsec:data_setup}

We use the composite CO survey of \citet{Dame+2001}, which traces molecular gas concentrated near the Galactic plane\footnote{\url{https://lweb.cfa.harvard.edu/rtdc/CO/CompositeSurveys/}}.
We select the region $|\ell|\leq90^{\circ}$, $|b|\leq5^{\circ}$, and $|v|\leq250\,{\rm km\,s^{-1}}$, and compare the latitude-integrated CO $\ell$--$v$ structure with the forward model.
Voxels not covered by the survey, or with missing data, are excluded using a data-coverage mask when needed.
Thus, the target data used in this section is not the full three-dimensional $T_{\rm b}(\ell,b,v)$ cube itself, but the latitude-integrated CO $\ell$--$v$ map constructed from it.

\begin{figure*}
\begin{center}
\includegraphics[width=0.95\textwidth]{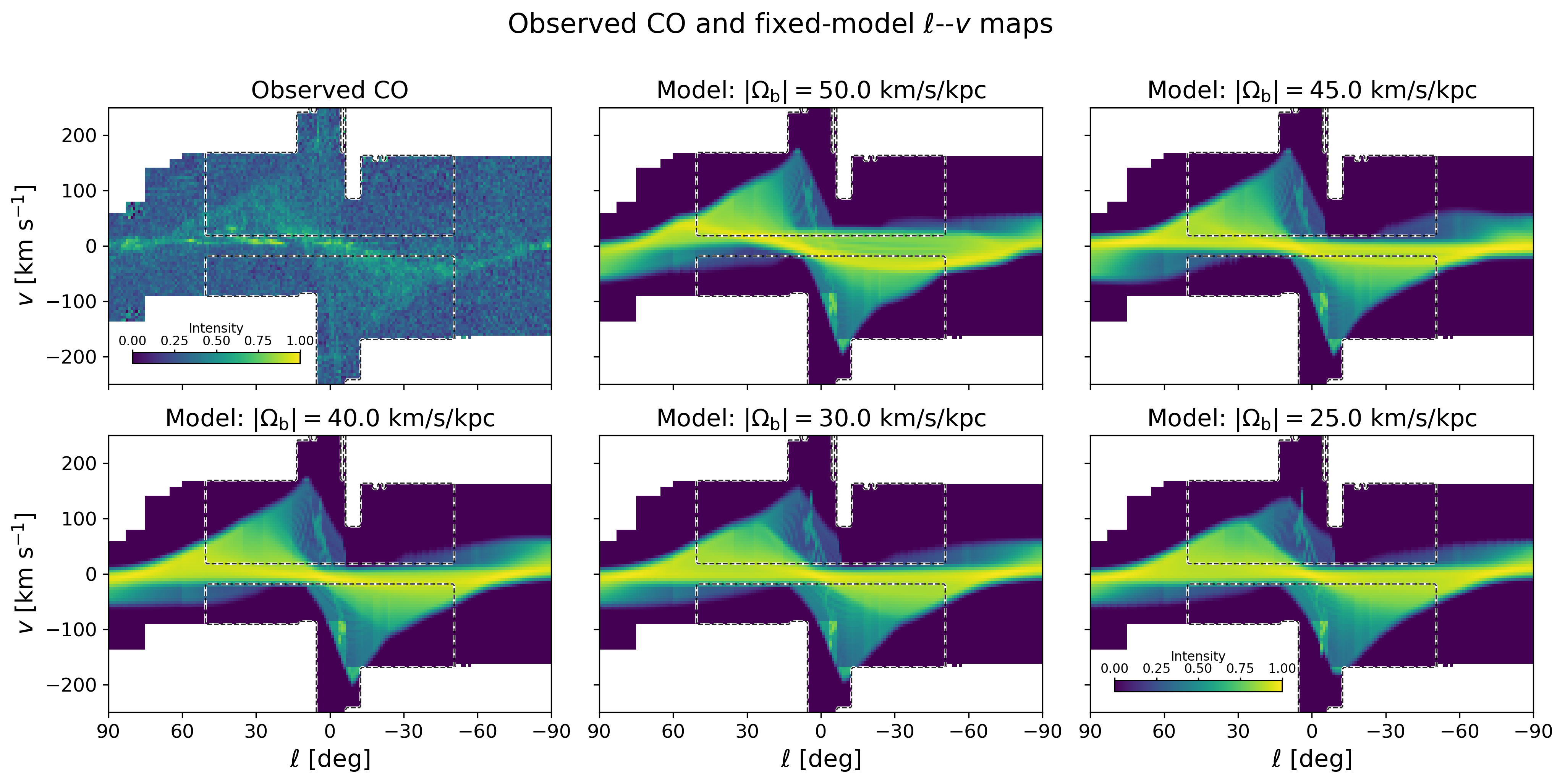}
\end{center}
\caption{
Observed CO $\ell$--$v$ map and fixed-model $\ell$--$v$ maps before optimization.
The upper-left panel shows the observed latitude-integrated CO map.
The other panels show forward-model maps for $|\Omega_{\rm b}|=50$, $45$, $40$, $30$, and $25\,{\rm km\,s^{-1}\,kpc^{-1}}$, with fixed $t_{\rm end}=0.30$ and $\phi_{\rm b}=25^{\circ}$.
The dashed contours show the fixed terminal-like fitting mask used in the observed-data fits; the same mask is applied to both the processed model and target maps.
\textbf{Alt text:} Observed CO longitude--velocity map compared with fixed model maps for several bar pattern speeds. Dashed contours show the terminal-like mask used to compare the processed observed and model maps.
}
\label{fig:data_fixed_compare}
\end{figure*}

The fitting model is the differentiable barred-gas flow model described in Section~\ref{sec:methods}.
Following the fiducial setup adopted in the independent-mock validation, we use a polar grid with $R=0.5$--$10\,{\rm kpc}$, adopt $\phi_{\rm b}=25^{\circ}$ unless otherwise stated, and vary $\Omega_{\rm b}$ and $t_{\rm end}$ over the ranges specified below.
The simulated gas is projected into observable $\ell$--$v$ space using the same soft-binning prescription as in Section~\ref{subsec:lv_projection}.
Because the model does not include tracer-specific chemistry or full line radiative transfer, we treat the projected surface density as an effective emission weight and compare processed morphology rather than absolute brightness.
For the observed CO comparison, we add a small intensity floor, of order $10^{-4}\,{\rm K}$, before applying the processing operator ${\cal P}$ defined in Section~\ref{subsec:objective_optimization}.
As above, the loss is interpreted as a morphology-based mismatch rather than a fit to the absolute CO brightness.
For the terminal-like fits, we define a fixed fitting mask in the $\ell$--$v$ plane that selects the high-velocity envelope of the observed CO emission.
The mask is defined by the longitude--velocity boundaries shown by the dashed contours in Figure~\ref{fig:data_fixed_compare}, and is held fixed for all models, optimization iterations, and parameter scans in a given comparison.
The same mask is applied to both the processed model and target maps before evaluating $\mathcal{L}$.

Before performing the gradient-based fit and parameter scans, we first inspect how the unoptimized forward models compare with the observed CO $\ell$--$v$ map.
Figure~\ref{fig:data_fixed_compare} shows the observed CO map and fixed-model maps for several pattern speeds,
$\Omega_{\rm b}=-50$, $-45$, $-40$, $-30$, and $-25\,{\rm km\,s^{-1}\,kpc^{-1}}$, at fixed $t_{\rm end}=0.30$ and $\phi_{\rm b}=25^{\circ}$.
The dashed contours indicate the terminal-like fitting region used below.
The fixed models do not reproduce the detailed CO morphology, including local gas, spiral-arm emission, near-center structure, and tracer-dependent small-scale features.
However, they show how changing $\Omega_{\rm b}$ shifts the broad high-velocity envelope and tilted emission in $\ell$--$v$ space.

\begin{figure*}
\begin{center}
\includegraphics[width=0.95\textwidth]{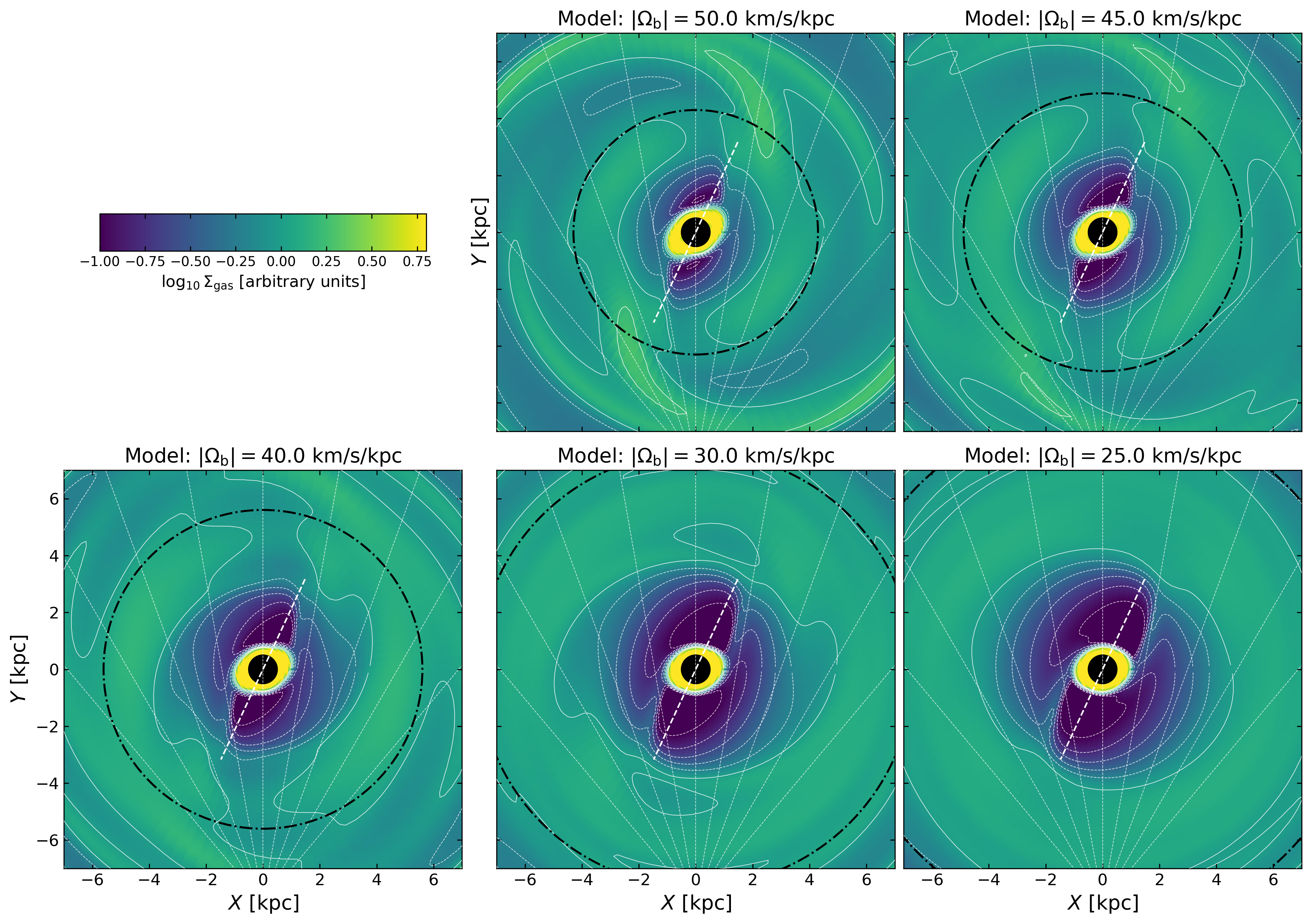}
\end{center}
\caption{
Face-on gas surface-density maps corresponding to the five fixed-pattern-speed models in Figure~\ref{fig:data_fixed_compare}.
All models use $t_{\rm end}=0.30$ and $\phi_{\rm b}=25^{\circ}$.
Colors and white contours show $\log_{10}\Sigma_{\rm gas}$ in arbitrary units.
The white dashed line marks the bar major axis, while the thin white dotted lines mark Galactic longitudes over $-50^{\circ}\leq\ell\leq50^{\circ}$ at $10^{\circ}$ intervals.
The black dot--dashed circles mark the corotation radii, $R_{\rm CR}$, defined by $\Omega_{\rm circ}(R_{\rm CR})=|\Omega_{\rm b}|$ using the axisymmetrized gravitational potential.
For $|\Omega_{\rm b}|=50$, $45$, $40$, $30$, and $25\,{\rm km\,s^{-1}\,kpc^{-1}}$, the corresponding corotation radii are $R_{\rm CR}=4.30$, $4.89$, $5.61$, $7.65$, and $9.15\,{\rm kpc}$, respectively.
The central black region corresponds to the inner radial boundary of the hydrodynamical domain at $R=0.5\,{\rm kpc}$.
As $|\Omega_{\rm b}|$ decreases, the bar-aligned low-density region broadens, while the dust-lane-like ridges bordering this region and the inner-ring-like structures near the bar ends change their positions, orientations, and radial extents.
\textbf{Alt text:}
Face-on gas surface-density maps for five bar pattern speeds. The nominal corotation radius moves outward as the pattern speed decreases, while the bar-aligned low-density region and surrounding dense gas ridges change in shape and extent.
}
\label{fig:data_fixed_faceon}
\end{figure*}

To connect the changes in the $\ell$--$v$ morphology with the underlying gas response, Figure~\ref{fig:data_fixed_faceon} shows the corresponding face-on gas surface-density maps.
The maps show a compact near-nuclear gas concentration, elongated dense ridges associated with dust-lane flows, and curved ridges near the bar ends that resemble parts of an inner ring. Such dust-lane shocks and nuclear rings are commonly produced in hydrodynamical models of barred galaxies \citep[e.g.,][]{Athanassoula1992b,ReganTeuben2003,Li+2015}.
The strongest pattern-speed dependence is seen in the bar-scale structures. 
As $|\Omega_{\rm b}|$ decreases, the corotation radius moves outward, the low-density region aligned with the bar broadens, and the dust-lane-like and bar-end ridges change their positions, orientations, and radial extents. 
The near-nuclear structure changes less strongly, although its interpretation is limited by the inner boundary of the hydrodynamical domain at $R=0.5\,{\rm kpc}$. 
Changes in the face-on positions and orientations of the ridges alter the Galactic longitudes at which they contribute most strongly to the line-of-sight projection, while changes in their streaming motions alter their line-of-sight velocities.
The corresponding $\ell$--$v$ ridges therefore shift in both longitude and velocity, changing the high-velocity envelope and forbidden-velocity emission \citep[e.g.,][]{Bissantz+2003,Sormani+2015c,Li+2016,Li+2022,Baba2025b}.
Figures~\ref{fig:data_fixed_compare} and \ref{fig:data_fixed_faceon} thus connect the pattern-speed dependence of the dust-lane and bar-end gas structures to that of the projected $\ell$--$v$ morphology.

\subsection{Representative gradient-based fit to the observed CO data}
\label{subsec:data_loss_consistency}

\begin{figure*}
\begin{center}
\includegraphics[width=0.95\textwidth]{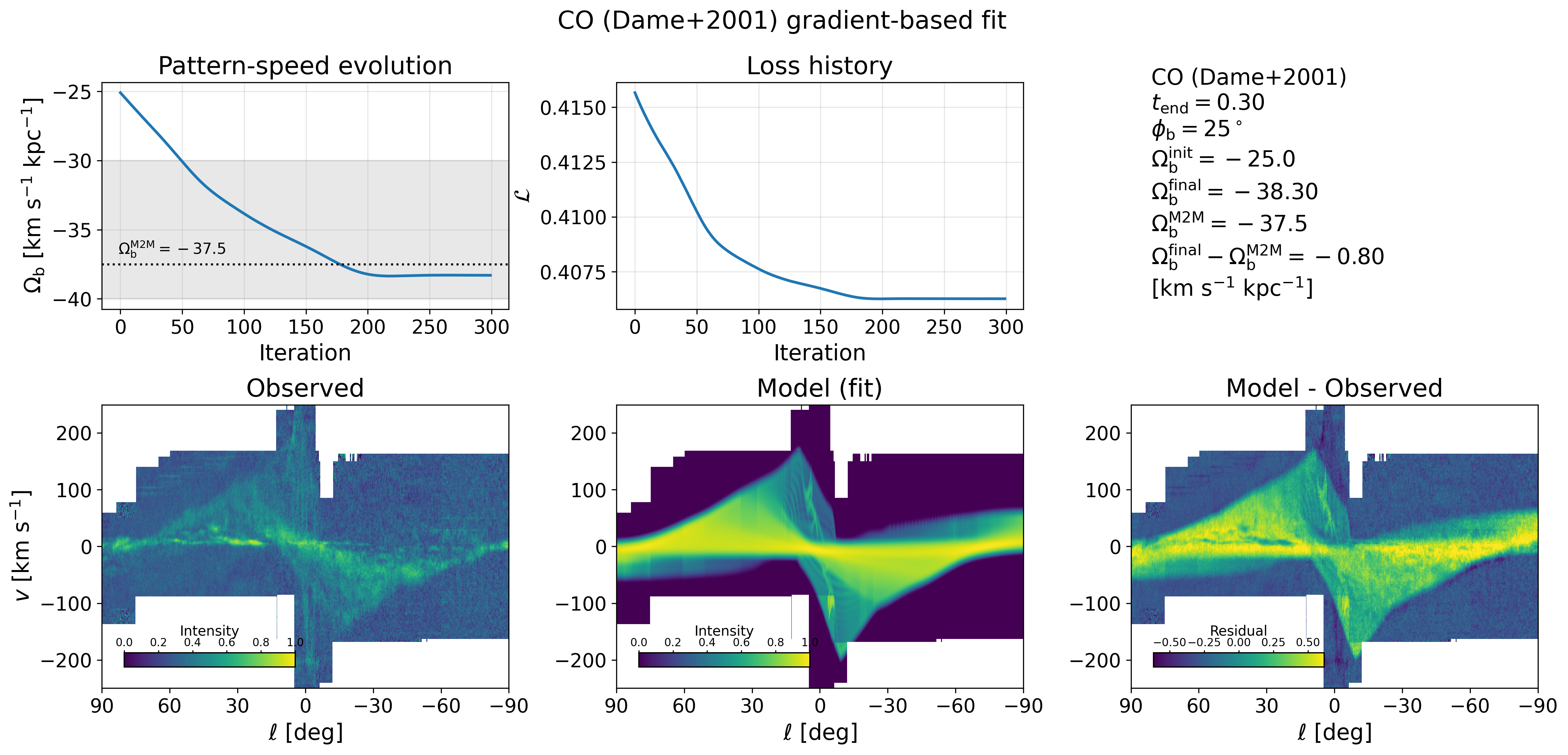}
\end{center}
\caption{
Representative gradient-based fit to the latitude-integrated CO $\ell$--$v$ map constructed from the Dame et al. CO cube.
The fit uses fixed $t_{\rm end}=0.30$ and $\phi_{\rm b}=25^{\circ}$, the terminal-like mask, and the initial value $\Omega_{\rm b}^{\rm init}=-25\,{\rm km\,s^{-1}\,kpc^{-1}}$.
The upper panels show the evolution of $\Omega_{\rm b}$ and the cosine-distance loss $\mathcal{L}$ during the optimization, together with a summary of the adopted and recovered values.
The dotted reference line in the pattern-speed panel marks the M2M bar-model pattern speed, $\Omega_{\rm b}^{\rm M2M}=-37.5\,{\rm km\,s^{-1}\,kpc^{-1}}$, adopted in the analytic barred potential of \citet{Sormani+2022agama} based on the M2M model of \citet{Portail+2017}.
The lower panels compare the observed processed $\ell$--$v$ map, the optimized model map, and the residual.
The fitted pattern speed reaches $\Omega_{\rm b}^{\rm final}=-38.30\,{\rm km\,s^{-1}\,kpc^{-1}}$ for this adopted response time, viewing angle, and mask.
This value should be interpreted as a diagnostic optimized point, not as a unique measurement of the Milky Way bar pattern speed.
Coherent residuals remain because the simplified model does not include tracer-specific CO emissivity, spiral arms, local gas, feedback, or full radiative transfer.
\textbf{Alt text:} Optimization history and final map comparison for a fit to the observed CO longitude--velocity map. The fitted pattern speed reaches about $-38\,{\rm km\,s^{-1}\,kpc^{-1}}$, but coherent residuals remain between the observed CO map and the simplified barred-gas model.
}
\label{fig:data_single_fit_co}
\end{figure*}

We first perform a representative gradient-based fit to the observed CO $\ell$--$v$ map using the same fiducial response time as in the mock validation, $t_{\rm end}=0.30$, and fixing the viewing angle to $\phi_{\rm b}=25^{\circ}$.
Figure~\ref{fig:data_single_fit_co} shows a fit initialized at $\Omega_{\rm b}^{\rm init}=-25\,{\rm km\,s^{-1}\,kpc^{-1}}$.
In the optimization-history panels, the dotted reference line marks the M2M bar-model pattern speed, $\Omega_{\rm b}^{\rm M2M}=-37.5\,{\rm km\,s^{-1}\,kpc^{-1}}$, adopted by \citet{Sormani+2022agama} for their analytic representation of the M2M barred Milky Way model of \citet{Portail+2017}.
The viewing angle is fixed to the fiducial value $\phi_{\rm b}=25^{\circ}$, representative of the current stellar-dynamical range, $\phi_{\rm b}\simeq25^{\circ}\pm10^{\circ}$ \citep{HuntVasiliev2025}.
The M2M pattern speed is shown only for comparison and is not used as a prior or constraint in the optimization.

The cosine-distance loss decreases during the optimization, and the fitted pattern speed moves toward larger $|\Omega_{\rm b}|$, reaching
$\Omega_{\rm b}^{\rm final}\simeq -38.30\,{\rm km\,s^{-1}\,kpc^{-1}}$.
This value is close to the M2M bar-model value, with $\Omega_{\rm b}^{\rm final}-\Omega_{\rm b}^{\rm M2M}\simeq -0.80\,{\rm km\,s^{-1}\,kpc^{-1}}$.
However, this single optimized value should be interpreted only as a diagnostic point.
In Section~\ref{subsec:data_systematics}, we examine how the result depends on nuisance choices such as the response time and viewing angle using explicit loss-map scans.

Physically, the optimization mainly tests whether the model can place the large-scale bar-driven high-velocity structures at approximately the observed longitudes and velocities.
The optimized map reproduces part of the global velocity envelope and tilted emission seen in the processed CO map, while coherent residuals remain in local low-velocity gas, spiral-arm emission, near-center gas, and tracer-dependent small-scale structures.
Because the adopted loss is a cosine-distance loss, the optimization emphasizes the orientation of the processed map in masked pixel space rather than the absolute CO brightness scale.
Thus, a lower loss should be interpreted as a better match to coherent large-scale kinematic morphology, not as a reconstruction of the full gas flow at every position in the Milky Way.

Because the simplified model has substantial systematic errors and $\mathcal{L}$ is a morphology-based objective function rather than a calibrated likelihood (Section~\ref{subsec:objective_optimization}), we do not use a single gradient-based trajectory to define a unique best-fitting pattern speed.
Instead, in the next subsection we examine explicit loss-map scans in $(t_{\rm end},\Omega_{\rm b})$ and $(\phi_{\rm b},\Omega_{\rm b})$.
These scans provide a more direct way to identify broad low-loss regions and to assess the systematic dependence on nuisance parameters.

\subsection{Dependence on gas response time and viewing angle}
\label{subsec:data_systematics}

The representative optimization in Figure~\ref{fig:data_single_fit_co} gives one diagnostic point in the observed-data loss landscape.
However, the optimized pattern speed can depend on nuisance choices such as the gas response time, viewing angle, fitting mask, map processing, and simplified emission model.
We therefore examine explicit loss-map scans rather than using a single optimization trajectory as a unique estimate of $\Omega_{\rm b}$.

\begin{figure*}
\begin{center}
\includegraphics[width=0.95\textwidth]{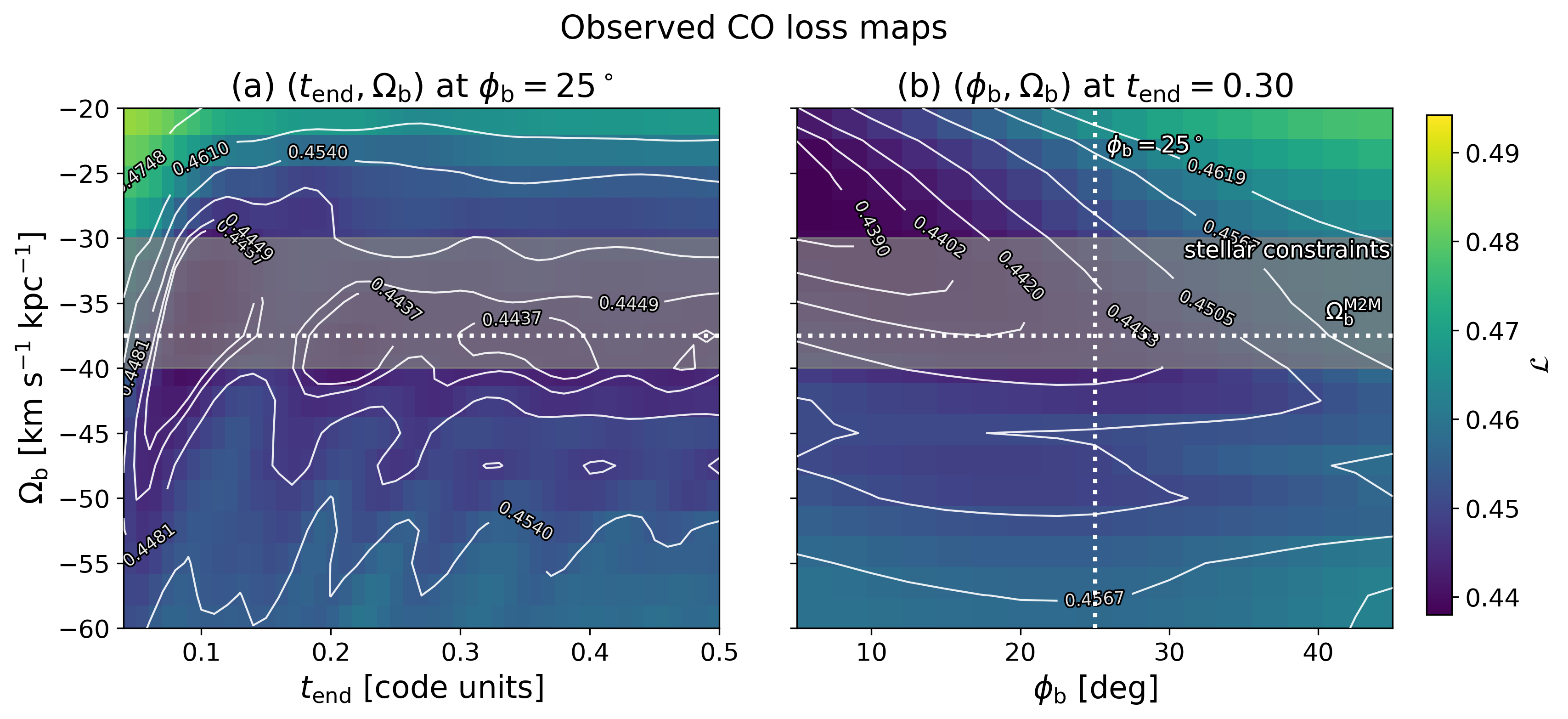}
\end{center}
\caption{
Observed CO loss maps.
Panel (a) shows the CO loss in the $(t_{\rm end},\Omega_{\rm b})$ plane at fixed $\phi_{\rm b}=25^{\circ}$.
Panel (b) shows the CO loss in the $(\phi_{\rm b},\Omega_{\rm b})$ plane at fixed $t_{\rm end}=0.30$.
In both panels, the plotted quantity is the cosine-distance loss $\mathcal{L}$ computed from the processed observed CO $\ell$--$v$ map using the terminal-like mask.
The shaded horizontal band marks the broad range $|\Omega_{\rm b}|=30$--$40\,{\rm km\,s^{-1}\,kpc^{-1}}$ representative of recent stellar-dynamical constraints.
The horizontal dotted reference line marks the M2M bar-model pattern speed, $\Omega_{\rm b}^{\rm M2M}=-37.5\,{\rm km\,s^{-1}\,kpc^{-1}}$, and the vertical dotted reference line in panel (b) marks the fiducial viewing angle, $\phi_{\rm b}=25^{\circ}$, representative of the current stellar-dynamical range.
The CO loss shows broad low-loss regions that overlap the stellar-dynamical range for plausible response times and viewing angles.
The minima should not be interpreted as formal maximum-likelihood estimates, because the loss is a morphology-based objective function and the model errors are not described by a full statistical noise model.
\textbf{Alt text:} Observed CO loss maps showing how the morphology-based loss depends on bar pattern speed, gas response time, and viewing angle. Broad low-loss regions overlap the stellar-dynamical pattern-speed range of $|\Omega_{\rm b}|=30$--$40\,{\rm km\,s^{-1}\,kpc^{-1}}$.
}
\label{fig:data_phi_omega_lossmaps}
\end{figure*}

We next examine the loss landscape explicitly.
Figure~\ref{fig:data_phi_omega_lossmaps}a shows the two-dimensional CO loss map in the $(t_{\rm end},\Omega_{\rm b})$ plane at fixed $\phi_{\rm b}=25^{\circ}$, while Figure~\ref{fig:data_phi_omega_lossmaps}b shows the loss map in the $(\phi_{\rm b},\Omega_{\rm b})$ plane at fixed $t_{\rm end}=0.30$.
In both panels, the loss is computed from the observed CO $\ell$--$v$ map using the terminal-like mask.

In the $(t_{\rm end},\Omega_{\rm b})$ plane, the low-loss region is broad and structured rather than a single isolated minimum.
At early response times, the preferred pattern speed changes rapidly with $t_{\rm end}$.
For $t_{\rm end}\gtrsim0.1$, low-loss regions extend over a broad range around
$\Omega_{\rm b}\simeq -30$ to $-40\,{\rm km\,s^{-1}\,kpc^{-1}}$, with additional branches at larger $|\Omega_{\rm b}|$ for some response times.
At the fiducial response time $t_{\rm end}=0.30$, the low-loss region includes
$\Omega_{\rm b}\simeq -35$ to $-40\,{\rm km\,s^{-1}\,kpc^{-1}}$, consistent with the representative gradient-based fit in Figure~\ref{fig:data_single_fit_co}.
This behavior indicates that the inferred pattern speed depends on the response stage of the gas flow.
Thus, $t_{\rm end}$ should be treated as a nuisance parameter controlling the instantaneous gas response, rather than as a directly measured physical age of the bar.

The $(\phi_{\rm b},\Omega_{\rm b})$ map shows a clear dependence on viewing geometry.
The scan covers $\phi_{\rm b}=5^{\circ}$--$45^{\circ}$, which includes and extends beyond the plausible viewing-angle range of the Milky Way bar.
At small viewing angles, the low-loss region is found near relatively slow pattern speeds, whereas at larger viewing angles it shifts toward larger $|\Omega_{\rm b}|$.
This trend shows a partial degeneracy between $\Omega_{\rm b}$ and $\phi_{\rm b}$ in the observed CO $\ell$--$v$ comparison.
Within the plausible stellar-dynamical viewing-angle range, $\phi_{\rm b}\simeq25^{\circ}\pm10^{\circ}$ \citep{Bland-HawthornGerhard2016,HuntVasiliev2025}, the low-loss region overlaps moderate-speed barred-flow solutions with
$|\Omega_{\rm b}|\simeq30$--$40\,{\rm km\,s^{-1}\,kpc^{-1}}$.
Thus, over the plausible viewing-angle range, the viewing geometry introduces a systematic uncertainty of order several ${\rm km\,s^{-1}\,kpc^{-1}}$ in the pattern-speed value inferred from the CO morphology.

The low-loss regions in both panels overlap the broad stellar-dynamical range $|\Omega_{\rm b}|\simeq30$--$40\,{\rm km\,s^{-1}\,kpc^{-1}}$ \citep{HuntVasiliev2025}.
The M2M bar-model pattern speed, $\Omega_{\rm b}^{\rm M2M}=-37.5\,{\rm km\,s^{-1}\,kpc^{-1}}$, and the fiducial viewing angle adopted in this paper, $\phi_{\rm b}=25^{\circ}$, lie within or close to these low-loss regions.
However, because the location and width of the low-loss region depend on the adopted gas response time, viewing angle, and simplified emission model, we interpret this result as a broad consistency check rather than a precision measurement of $\Omega_{\rm b}$, and discuss its implications in Section~\ref{sec:discussion}.

The present analysis focuses on CO because it gives a clearer large-scale match in the current simplified model.
In principle, the same differentiable framework can be applied to H\,{\sc i} and other gas tracers, but such applications require tracer-specific treatment of emissivity, optical-depth effects, survey selection, and vertical gas structure.
We leave a full multi-tracer extension to future work.

\section{Summary and Discussion}
\label{sec:discussion}

\subsection{Summary}
\label{subsec:summary_main}

We have developed a differentiable hydrodynamical framework for comparing barred-gas models of the Milky Way directly with observed $\ell$--$v$ data.
Previous gas-dynamical studies have compared simulations with observed gas features, terminal-velocity curves, or discrete sets of models \citep[e.g.,][]{Fux1999,Bissantz+2003,SormaniMagorrian2015,Sormani+2015c,Li+2016,Li+2022}, whereas here we formulate the comparison as a differentiable forward-modeling problem.
To our knowledge, this is the first application of differentiable hydrodynamical modeling to Milky Way gas dynamics.
The model evolves an isothermal gas disk in a fixed barred potential, projects the gas distribution into observable $\ell$--$v$ space with a differentiable soft-binning operator, and evaluates the mismatch between the model and target maps.
In the present implementation, this mismatch is measured with a cosine-distance loss applied to processed and masked $\ell$--$v$ maps, so that the comparison focuses on large-scale morphology rather than on the absolute intensity scale.
Because the full forward model is differentiable, the loss gradient can be propagated back to dynamical parameters such as the bar pattern speed, $\Omega_{\rm b}$.

We validated the method with self-consistency hydrodynamical mocks and with an independent hydrodynamical mock generated by a different numerical method and more complete interstellar-medium physics.
The self-consistency tests recover the input pattern speed accurately.
The independent-mock test identifies a broad low-loss region around the input value, although the optimization can converge to secondary branches depending on the initial value, fitting mask, gas response time, viewing angle, and radial domain.
These tests show that the method captures coherent large-scale bar-driven kinematic structure in $\ell$--$v$ space, rather than requiring an exact match to the detailed gas morphology or absolute emission amplitude.

We then applied the same framework to the observed CO data in the inner Milky Way.
The observed-data comparison should be interpreted as a match to large-scale bar-driven $\ell$--$v$ morphology, not as a full reconstruction of the Galactic gas flow, because the simplified model omits several tracer-dependent and small-scale gas structures discussed in Section~\ref{sec:data_application}.

For the fiducial setup with $t_{\rm end}=0.30$ and $\phi_{\rm b}=25^{\circ}$, a representative gradient-based fit initialized at $\Omega_{\rm b}^{\rm init}=-25\,{\rm km\,s^{-1}\,kpc^{-1}}$ reaches a moderate-speed solution near $\Omega_{\rm b}^{\rm final}\simeq -38.30\,{\rm km\,s^{-1}\,kpc^{-1}}$.
This optimized value is useful as a diagnostic example, but we do not use this single optimization run to define a unique best-fitting pattern speed.
Instead, our interpretation is based mainly on explicit loss maps, which show broad low-loss regions rather than a single isolated global minimum.
Within the plausible stellar-dynamical viewing-angle range $\phi_{\rm b}\simeq25^{\circ}\pm10^{\circ}$, the CO loss maps show broad low-loss regions that overlap moderate-speed barred-flow solutions with $|\Omega_{\rm b}|\sim30$--$40\,{\rm km\,s^{-1}\,kpc^{-1}}$.
This low-loss region overlaps the broad stellar-dynamical constraints on the Milky Way bar \citep{HuntVasiliev2025}.
It also includes the pattern speed adopted in the M2M-based barred potential used in this work, $\Omega_{\rm b}^{\rm M2M}=-37.5\,{\rm km\,s^{-1}\,kpc^{-1}}$.
Thus, the observed CO $\ell$--$v$ morphology is broadly consistent with a moderate-speed barred Milky Way model, including the M2M-based reference model, within the systematic uncertainties explored here.

\subsection{Relation to previous studies}
\label{subsec:discussion_physical}

These results are consistent with previous gas-dynamical models of the Milky Way bar.
Earlier work showed that large-scale $\ell$--$v$ features provide strong constraints on barred-gas models, and that automatic quantitative searches are needed because different features can favor different model parameters \citep{SormaniMagorrian2015,Sormani+2015c}.
\citet{Li+2016} showed that a low-pattern-speed bar model with $|\Omega_{\rm b}|\simeq33\,{\rm km\,s^{-1}\,kpc^{-1}}$ can reproduce many observed gas features, and \citet{Li+2022} later favored $|\Omega_{\rm b}|=37.5$--$40\,{\rm km\,s^{-1}\,kpc^{-1}}$ using updated Milky Way potentials and additional gas-kinematic constraints.
The moderate-speed bar favored by the present observed-data comparison is consistent with this broad picture.

Our differentiable forward-modeling framework differs from these studies in methodology.
Instead of building a detailed best-fitting gas model from selected $\ell$--$v$ features, terminal velocities, and high-mass star-forming-region kinematics, we compare the model and data through a differentiable loss defined on the processed $\ell$--$v$ maps.
The use of a cosine-distance loss makes this comparison insensitive to an overall intensity scale and emphasizes the morphology of the processed maps.
This approach is complementary to detailed hydrodynamical modeling because it enables direct optimization and controlled exploration of parameter degeneracies, although the current implementation remains simplified.
The expected effects of additional gas physics on the predicted $\ell$--$v$ diagram are discussed in Section~\ref{subsec:discussion_ism_physical}. Given these model limitations, the present CO application should be viewed as a gas-kinematic consistency test rather than a final precision measurement of the Milky Way bar pattern speed.

This comparison with previous gas-dynamical studies also clarifies which parts of the $\ell$--$v$ structure carry most of the information on $\Omega_{\rm b}$ in our analysis.
The mock tests and observed-data loss maps indicate that the sensitivity to $\Omega_{\rm b}$ mainly comes from coherent large-scale structures, such as high-velocity envelopes, tilted emission ridges, and forbidden-velocity features, rather than from small-scale clumps.
This behavior is expected because changing $\Omega_{\rm b}$ changes the phase and strength of the bar-driven gas response, which in turn shifts the projected $\ell$--$v$ morphology.
At the same time, similar changes can be partly compensated by changing the viewing angle $\phi_{\rm b}$ or the gas response time $t_{\rm end}$.
This explains why the observed-data comparison yields broad low-loss regions rather than a single sharply defined best-fitting pattern speed.

\subsection{Expected effects of additional gas physics on the $\ell$--$v$ diagram}
\label{subsec:discussion_ism_physical}

The present differentiable model assumes a two-dimensional, isothermal, non-self-gravitating gas disk. Additional gas physics can affect the width, contrast, fragmentation, and time variability of the predicted $\ell$--$v$ features. Even within isothermal models, changing the effective sound speed can alter the locations and shapes of shocks and the nuclear gas structure in a barred potential \citep{EnglmaierGerhard1997}. Simulations that include gas self-gravity and radiative heating and cooling show that a multi-phase gas disk develops dense clumps and filaments, and that the morphology of shocks and nuclear rings can differ substantially from that in non-self-gravitating isothermal models \citep[e.g.,][]{WadaKoda2001,Dobbs2023review}. Galactic simulations that also include star formation and stellar feedback produce both large-scale and clumpy structures in synthetic $\ell$--$v$ diagrams \citep{Baba+2010}. Recent self-consistent simulations with different gas treatments also show substantial model-to-model differences in the H~{\sc i} terminal-velocity curves, although these calculations do not isolate gas physics from changes in the formation and evolution of the bar \citep{Davis+2026}. Additional gas physics may therefore modify both the small-scale structure and the broad morphology of the predicted $\ell$--$v$ diagram.

The independent SPH mock in Section~\ref{subsec:sph_mock}, which is based on a three-dimensional SPH simulation with gas self-gravity, radiative cooling, star formation, and stellar feedback, provides a complementary test of the robustness of the pattern-speed inference.
Although its $\ell$--$v$ map is more clumpy and shows a broader emission envelope and more detailed high-velocity structure than the simplified fitting model, the inferred broad low-loss region includes the input pattern speed. Thus, at least in this mock test, the additional small-scale complexity does not erase the broad pattern-speed signal. However, this robustness is demonstrated only for the present mock test. Controlled comparisons under the same fixed barred potential will be required to quantify the effects of individual physical processes on the inferred pattern speed.

\subsection{Limitations and outlook}
\label{subsec:future_extensions}

Beyond the simplified gas physics discussed in Section~\ref{subsec:discussion_ism_physical}, the main limitations of the present implementation are the fixed gravitational potential and the morphological objective function.

In the present implementation, the gravitational potential is fixed to the stellar-dynamically constrained barred Milky Way model \citep{Portail+2017,Sormani+2022agama}, so we optimize only the pattern speed while keeping the bar mass distribution, scale lengths, disk, nuclear stellar component, and dark halo fixed.
In principle, the same differentiable framework can be extended to optimize additional parameters of the potential, such as the bar mass, scale length, shape, or the relative weights of the axisymmetric and non-axisymmetric components, which would allow us to infer the Milky Way potential and the gas response at the same time.
However, such an extension would also make the problem more degenerate, because similar $\ell$--$v$ structures could be produced by changing the gravitational potential, the CO emissivity, or the gas response time.
Future applications with a flexible potential will therefore require stronger priors and more realistic gas physics.

The second limitation is that the current objective function is morphological.
The cosine-distance loss is useful for this proof-of-concept study because it avoids an additional weighting hyperparameter and reduces sensitivity to the uncertain CO intensity scale.
However, it does not use the absolute brightness information and should not be interpreted as a statistical likelihood.
A future likelihood-based analysis would require a noise model for the observations and an effective model for the systematic mismatch, including missing spiral structure, local gas, tracer-dependent emissivity, optical-depth effects, and unresolved cloud-scale physics.

Because the present application optimizes mainly one physical parameter, explicit loss-map scans are still feasible and are useful for interpreting the result.
The advantage of the differentiable approach will become clearer once the model includes more parameters, such as the bar mass, viewing angle, gas response time, or emissivity parameters, for which exhaustive grid searches quickly become impractical.
The present study should therefore be regarded as a proof of concept for future multi-parameter differentiable inference of Milky Way gas dynamics.

A further extension is to use the gradient information itself as a diagnostic.
Because the forward model is differentiable, sensitivity maps can be constructed by combining the gradient of the cosine-distance loss with respect to the processed map, $\partial \mathcal{L}/\partial \hat{I}(\ell,v)$, with the response of the predicted map to the pattern speed, $\partial \hat{I}(\ell,v)/\partial \Omega_{\rm b}$.
Such maps would identify which $\ell$--$v$ regions, such as the terminal-velocity envelope, forbidden-velocity emission, tilted ridges, the 3-kpc-arm region, or near-center gas, carry the dominant pattern-speed information, and could clarify the origin of the partial degeneracy between $\Omega_{\rm b}$ and $\phi_{\rm b}$.

More broadly, this study is a first step toward fully forward-modeling Galactic gas line-intensity data in position--position--velocity (PPV) space, where PPV denotes Galactic longitude, latitude, and line-of-sight velocity $(\ell,b,v)$.
This goal is closely related to the bar-informed kinematic-distance framework of \citet{Baba2026a}, which provides a practical map-level reconstruction of the inner Milky Way.
Because kinematic-distance methods rely on an assumed streaming field, the present hydrodynamical forward model could supply that field, while the reconstructed gas maps could in turn provide stronger constraints on the forward model.
The observed CO data favor broad low-loss regions that overlap the stellar-dynamical constraints, despite the degeneracy between $\Omega_{\rm b}$ and $\phi_{\rm b}$, which suggests that differentiable gas dynamics can serve as a useful physical forward model for future Milky Way gas mapping and for joint inference of gas density and non-circular motions.

\section*{Funding}
This research was supported by the Japan Society for the Promotion of Science (JSPS) under Grant Numbers 21K03633, 21H00054, 22H01259, 24K07095, and 25H00394.

\section*{Data availability} 
The simulation snapshots and analysis code are available from the corresponding author upon reasonable request.

\begin{ack}
We sincerely thank the anonymous referee for their thoughtful and constructive comments, which helped improve the clarity and context of this paper.
We thank Takafumi Tsukui and Rimpei Chiba for helpful discussions.
The SPH simulation used to construct the independent mock data was carried out on Cray XD2000 (ATERUI-III) of the Center for Computational Astrophysics, National Astronomical Observatory of Japan (CfCA/NAOJ).
\end{ack}


\end{document}